# Inconsistent black hole kick estimates from gravitational-wave models

## Angela Borchers[1,2,*]  and Frank Ohme[1,2]

[1] Max Planck Institute for Gravitational Physics (Albert Einstein Institute), Callinstraße 38, D-30167 Hannover, Germany
[2] Leibniz Universität Hannover, 30167 Hannover, Germany

E-mail: angela.borchers.pascual@aei.mpg.de

**Abstract**
The accuracy of gravitational-wave (GW) models of compact binaries has traditionally been addressed by the mismatch between the model and numerical-relativity (NR) simulations. This is a measure of the overall agreement between the two waveforms. However, the largest modelling error typically appears in the strong-field merger regime and may affect subdominant signal harmonics more strongly. These inaccuracies are often not well characterised by the mismatch. We explore the use of a complementary, physically motivated tool to investigate the accuracy of GW harmonics in waveform models: the remnant's recoil, or kick velocity. Asymmetric binary mergers produce remnants with significant recoil, encoded by subtle imprints in the GW signal. The kick estimate is highly sensitive to the intrinsic inaccuracies of the modelled GW harmonics during the strongly relativistic merger regime. Here we investigate the accuracy of the higher harmonics in four state-of-the-art waveform models of binary black holes. We find that the SEOBNRv4HM_ROM, IMRPhenomHM, IMRPhenomXHM and NRHybSur3dq8 models are not consistent in their kick predictions. Our results enable us to identify regions in the parameter space where the models require further improvement and support the use of the kick estimate to investigate waveform systematics. We discuss how NR kick estimates could be used to calibrate waveform models further, proposing the first steps towards kick-based gravitational-wave tuning.

[*] Author to whom any correspondence should be addressed.

## 1. Introduction

Every gravitational-wave (GW) transient detection reported by the LIGO, Virgo and KAGRA (LVK) Collaborations to date is consistent with the expected signal radiated by merging compact binaries [1–4]. In order to extract the properties imprinted in the detected GW signals, Bayesian parameter estimation studies rely on the use of gravitational waveform models [5, 6]. These models need to be sufficiently accurate such that they do not cause any systematic biases when analysing the observed data. Systematic uncertainties of currently available models are in most cases below the statistical errors of the Advanced LIGO [7] and Advanced Virgo [8] detectors [9]. However, in light of the enhancement in the sensitivity of the current and next generation of GW detectors, the systematic errors of the models will have to be further reduced so that they are consistently below the instrumental ones. In particular, for the third-generation detector network, the mismatch errors of semi-analytical waveform models will need to be reduced by at least three orders of magnitude [10].

Having an accurate description of the GW higher harmonics (also named modes or multipoles) is relevant for several reasons. The inclusion of subdominant harmonics significantly improves the description of the signal, in particular, when characterising asymmetric binary systems. Indeed, the omission of higher harmonics in a waveform model can influence the search and interpretation of particular GW signals when the binary is oriented in specific configurations and can lead to systematic biases in the estimation of the source parameters [11–15]. The development of waveform models which include higher harmonics and the regular detection of GWs have recently allowed performing such analyses with observed GW signals of binary black holes (BBHs). The two events GW190412 [16] and GW190814 [17] observed by the LIGO and Virgo detectors, have provided strong evidence for the measurement of gravitational higher harmonics in the observed signals. Besides, as studied in [18], the inclusion of subdominant harmonics can reduce the uncertainty of parameter estimation results and can even shift posterior samples, as shown in the re-analysis of GW170729 [19] with the latest generation of phenomenological waveform models [20–26].

As detectors become more sensitive, GW observations will become more frequent and will include higher signal-to-noise ratios. If the waveform models employed in the analysis of such events are not reliable enough, we might mischaracterise the sources of these signals. Conversely, the existence of 'louder' GW events will allow us to observe more subtle physical effects in the signals, such as the remnant's kick. Asymmetric compact binaries radiate linear momentum through GWs causing, in turn, the recoil of the remnant object. This process is left imprinted in the emitted GW signal. Because of its astrophysically important consequences (see [27] for a review), there is a strong interest in inferring the remnant's kick velocity directly from GW signals. Several methods have been proposed to discuss the detectability of kicks in GW events [28–30]. Among the current GW observations, GW190814 shows the most informative kick estimate from the events of the second GW transient catalog (GWTC-2) [31], while GW200129_06 5458 shows support for a large kick velocity [32]. However, the kick's subtle signatures have not been precisely measured so far. If we want to observe these in the signal, we need waveform models that accurately capture such traces.

The accuracy of waveform models has traditionally been quantified by comparing the waveform model to numerical relativity (NR) or hybrid waveforms (see e.g. [33]). NR simulations yield the most accurate description of the GW signal and they are the only way to access the highly dynamic merger regime from first principles. The standard method of comparison is based on calculating a match between the model and an NR or hybrid waveform, based on a Wiener inner product. This is the same inner product that is used to assess the likelihood of whether the GW data contain a signal, by comparing the agreement between observational data and waveform templates. The match represents a notion of the angle between two signals (related to a distance between them) and is a standard quantity used in the waveform modelling community to test the quality of a model (see e.g. [34]).

A complementary tool to standard match calculations has recently been suggested in [35]. The authors propose an infinite set of constraints on compact binary coalescence waveforms, predicted by full, non-linear general relativity (GR). The set of constraints are the Bondi-van der Burg–Metzner–Sachs (BMS) balance laws, which are induced by the infinite-dimensional group of supertranslations [36], the natural extension of the four-dimensional group of translations defined at null infinity. The theory provides the opportunity to test waveform systematic errors in a new way. The balance laws can be particularly useful for regions of the parameter space where performing NR simulations might be more challenging or where these simulations are not abundant enough for the subsequent calibration of the approximate waveform models. Because these constraints come from exact GR, the balance laws are also attractive to quantify the accuracy of NR waveforms, which are often considered as a proxy of the exact waveform predicted by GR. The set of balance laws have already been applied to currently available waveform models to test their accuracy based on their angular momentum [37] and the gravitational memory estimates [38]. Besides, the Simulating eXtreme Spacetimes (SXS) Collaboration has recently employed the BMS balance laws to correct the extracted SXS waveforms by incorporating gravitational memory effects [39]. Apart from imposing accuracy requirements, the theory presented in [35] represents new tests of GR.

Although the expression of the radiated three-momentum flux by itself does not represent a BMS balance law, the calculation of the linear momentum flux (or the kick velocity) has recently been used to test the accuracy of the relative phase shifts between the GW harmonics of the state-of-the-art phenomenological waveform models [21, 25]. These studies have focused on the dependency of the kick magnitude on the symmetric mass ratio of the binary.

Using the recoil prediction to assess the waveform accuracy is attractive for three main reasons: (1) The kick estimate is a highly sensitive quantity [40–42]. The presence of small time and phase deviations that appear from incorrect modelling can result in significantly different kick predictions. (2) The kick builds up in the merger, the highly dynamic region that is most complex to model. Therefore, a correct kick estimate requires an accurate description of the merger. (3) The asymmetries of the system that lead to the kick are most completely characterised when using the dominant and higher harmonics. Thus, an accurate description of the subdominant harmonics is essential for a correct prediction of the kick velocity.

In this paper, we explore the use of the kick velocity as a diagnostic test for waveform models, by analysing the waveform estimates of the magnitude and orientation of the kick velocity over the parameter space. Precessing binaries lead to the largest recoil velocities [43] and are therefore observationally easier to accesss. However, only one of the available precessing waveform models includes the mode asymmetries that generate out-of-plane kicks for precessing binaries: the NR Surrogate (NR Sur) model. This model only covers part of the parameter space relevant for LVK data analysis. Kicks in nonprecessing binaries, on the other hand, can be predicted by any of the nonprecessing models. For this reason, in this work, we focus on nonprecessing systems as a first step to understanding the accuracy of GW harmonics through remnant kicks.

We evaluate the accuracy of four waveform models which are used in current GW data analysis studies. We also analyse the performance of the NR Sur fit [44], recently used to make the first measurement of a large kick velocity in an observed GW signal [32]. By exploiting the features of the kick, we create a set of diagnostic tests that can be applied to any waveform model to identify modelling inaccuracies over the parameter space.

After a description of the methodology of our work in section 2, we use the kick velocity to evaluate the accuracy of several gravitational waveforms and analyse their harmonic contributions to the kick in section 3. In section 4 we further explore the applications of the kick in the context of waveform modelling by addressing the first steps towards kick-based GW tuning. Finally, we summarise our results in section 5.

## 2. Method

### 2.1. Waveform models

The GW signal radiated by a BBH coalescence is uniquely determined by a number of physical parameters that characterise the binary. Astrophysical quasi-circular BBHs are described by the two individual masses $m_i$ and the individual spin vectors. It is common to use the dimensionless spin parameter $\vec{\chi}_i = \vec{S}_i / m_i^2$ for the spin components. BBHs on quasi-circular orbits are thus characterised by eight intrinsic parameters:

$$\lambda = \{m_1, m_2, \vec{\chi}_1, \vec{\chi}_2\}. \tag{1}$$

The total mass of the binary is $M = m_1 + m_2$ and the symmetric mass ratio is defined as $\eta = m_1 m_2 / M^2$. Throughout this work we use geometric units, $G = c = 1$. Apart from the intrinsic parameters, there are seven extrinsic parameters: the spherical angles $\{\theta, \varphi\}$, which describe the orientation of the binary, the luminosity distance $d_L$, the coalescence time $t_c$, the declination and right ascension $\{\vartheta, \phi\}$, and the polarization angle $\psi$.

In NR, it is common to use the Newman–Penrose scalar, $\Psi_4$, which is derived from the Weyl tensor and encodes the outgoing gravitational radiation as $\Psi_4 = -\ddot{h}$, where $h := h_+ - i h_\times$. The behaviour of a quantity under rotations is expressed by the spin weight. Because the Weyl tensor component has spin weight $s = -2$, the GW strain can be expanded in a basis of spin-weighted spherical harmonics (SWSH) [45],

$$\lim_{r \to \infty} rh = \sum_{\ell=2}^{\infty} \sum_{m=-\ell}^{\ell} h_{\ell,m}(t, \lambda) \, ^{-2}Y_{\ell,m}(\theta, \varphi). \tag{2}$$

Here $h_{\ell,m}(t, \lambda)$ are the GW spherical harmonics associated to the multipole moments. The (2,2) spherical harmonic is the quadrupolar term, while those associated with higher multipole moments are referred to as higher harmonics. The $h_{\ell,m}(t, \vec{\lambda})$ depend on the time and the intrinsic physical properties of the source, $\vec{\lambda}$. The orientation of the source with respect to the observer is encoded in the spherical harmonic basis functions of spin weight $-2$, $\, ^{-2}Y_{\ell,m}(\theta, \varphi)$. The coalescence phase $\varphi_c$ is sometimes included as an extrinsic parameter of the binary, and is degenerate with the azimuthal angle $\varphi$ in nonprecessing systems.

In a nonprecessing binary, the spins of the individual black holes are parallel to the direction of the orbital angular momentum, and the evolution of the binary takes place in the plane perpendicular to the orbital angular momentum. In such case, there exists an equatorial symmetry between the $(\ell, m)$ and $(\ell, -m)$ modes, given by the relation:

$$h_{\ell,m}(t) = (-1)^\ell h^*_{\ell,-m}(t),\tag{3}$$

where $*$ refers to the complex conjugation.

**Table 1.** Gravitational waveform models used in our study. The second column indicates the higher harmonics included in each of the models.

| Waveform model | Multipoles $(\ell, |m|)$ |
|---|---|
| IMRPhenomHM | $(2,2), (2,1), (3,3), (3,2), (4,4), (4,3)$ |
| IMRPhenomXHM | $(2,2), (2,1), (3,3), (3,2), (4,4)$ |
| SEOBNRv4HM_ROM | $(2,2), (2,1), (3,3), (4,4), (5,5)$ |
| NRHybSur3dq8 | $\ell \leqslant 4, (5,5)$ but not $(4,0), (4,1)$ |

Although NR simulations determine the closest description of the real set of $h_{\ell,m}$ harmonics, for GW data analysis purposes, waveforms need to be fast to evaluate, and NR simulations are in this context computationally excessively expensive. For this reason, different modelling strategies have been developed, leading to the establishment of three main waveform families: the NR calibrated effective-one-body (EOBNR) [46–51], the phenomenological (Phenom) [20, 24, 52–59] and the NR Sur description [60–62]. These models characterise the full inspiral-merger-ringdown signal over the parameter space, and their development has been based on combining analytic and numerical methods appropriate for each of the different phases of the binary's evolution.

The inspiral and merger phases can be partly described by the EOB formalism, which maps the two-body problem to that of a test particle in an effective metric (see [63] for a review) and free coefficients are calibrated to NR data. On the other hand, Phenom models are based on employing hybrid waveforms that connect an analytical inspiral description with NR data for the late inspiral, merger and ringdown, which are then described by a phenomenological fit. The ringdown phase of the remnant black hole is characterised by the emission of quasinormal modes [64], mathematically described in terms of exponentially damped oscillations. In contrast, the NR Sur family is based on a reduced order method interpolation of the NR simulations over the parameter space, built to cover a larger region of the parameter space than that covered by the NR waveforms.

In our work, we analyse the kick predictions of four nonprecessing waveform models that include higher harmonics: SEOBNRv4HM_ROM [65], two phenomenological models, namely, IMRPhenomHM [57] and the more recent IMRPhenomXHM [21, 22], and NRHybSur3dq8 [61]. Table 1 indicates which gravitational multipoles are included in each waveform model.

## 2.2. Linear momentum flux

Because asymmetric BBHs radiate linear momentum through GWs, the remnant black hole acquires a kick velocity. This can be mathematically described from the linear momentum flux radiated by the binary. Since the linear momentum of the system is initially zero, the momentum of the remnant is equal to the opposite of the three-momentum flux carried by the radiated GWs, and is given by:

$$P_i = -\lim_{r \to \infty} \frac{r^2}{16\pi} \int_{-\infty}^{\infty} dt \oint d\Omega \, \hat{x}_i(\theta, \varphi) \, |\dot{h}|^2,\tag{4}$$

where $\mathrm{d}\Omega = \sin\theta\, \mathrm{d}\theta\, \mathrm{d}\varphi$ and $\hat{x}_i = (\sin\theta\cos\varphi, \sin\theta\sin\varphi, \cos\theta)$ is the unit vector expressed in the spherical harmonic basis. Because the asymmetries that cause the kick are encoded in the GW signal, the momentum of the remnant black hole is entirely determined by the waveform $h$. In asymmetric BBH systems, gravitational higher harmonics are particularly loud during merger. For this reason, we express the radiated momentum in terms of the dominant and higher multipoles by decomposing the gravitational radiation on a basis of SWSH.

We choose the $z$ axis along the orbital angular momentum of the binary. In a nonprecessing system, the binary orbits in a fixed plane, that we choose as the $x$–$y$ plane. In this case, the $z$-component of the momentum vanishes and the kick takes place in the orbital plane. The momentum can then be expressed as a combination of the two planar coordinates, $\vec{P} := P_x + iP_y$. After expressing the unit-vector in terms of the SWSH and integrating over the 2-sphere, one can show that the components of the momentum are given by [66]:

$$\vec{P} = -\frac{1}{8\pi}\int_{-\infty}^{\infty}\mathrm{d}t\sum_{\ell=2}^{\infty}\sum_{m=-\ell}^{\ell}\dot{h}_{\ell,m}(a_{\ell,m}\dot{h}_{\ell,m+1}^{*} + b_{\ell,-m}\dot{h}_{\ell-1,m+1}^{*} - b_{\ell+1,m+1}\dot{h}_{\ell+1,m+1}^{*}),\tag{5}$$

where the coefficients $a_{\ell,m}$ and $b_{\ell,m}$ read:

$$a_{\ell,m} = \frac{\sqrt{(\ell-m)(\ell+m+1)}}{\ell(\ell+1)},\tag{6}$$

$$b_{\ell,m} = \frac{1}{2\ell}\sqrt{\frac{(\ell-2)(\ell+2)(\ell+m)(\ell+m-1)}{(2\ell-1)(2\ell+1)}}.\tag{7}$$

The kick velocity will then be given by $\vec{v}_f = -\vec{P}/M_f$, where $M_f$ is the mass of the remnant black hole.

### 2.3. Implementation

We have implemented the expression of the momentum (5) for the mentioned gravitational waveform models. We use the LIGO Algorithm Library (LALSuite) software [67] to obtain individual higher harmonics of the GW signal. It is physically meaningful to compute the momentum flux in the time domain. However, for data analysis purposes the SEOB-NRv4HM_ROM and Phenom models that we employ in our study have been developed in the frequency domain. For this reason, we inverse Fourier transform the GW signal after obtaining the individual spherical harmonics in the frequency domain. We use $f_{\min} = 10$ Hz for the lower frequency cut-off and because the kick velocity does not depend on the total mass of the binary, we fix the total mass to $M = 50 M_{\odot}$. For a nonspinning equal-mass binary this corresponds to 63 orbits before merger. To calculate the remnant's final mass, we use waveform specific fitting functions which are available in LALSuite[3]. The final mass is internally computed from the radiated energy and is scaled by the original total mass. For details on how the radiated energy is calculated we refer the reader to the paper references.

We compare the kick estimates with the predictions of a set of NR waveforms. We have used SXS waveforms [68] from the LVCNR Waveform Catalog [69] with the highest resolution available. These include all subdominant harmonics up to $\ell = 8$, with $|m| \leqslant \ell$. The NR data come with a metadata file which includes a *coordinate* recoil velocity estimation that is calculated based on the trajectory $\vec{x}(t)$ of the coordinate centre of the apparent horizon of the remnant. Although this value is close to the one computed from the momentum flux integral, they might not be necessarily the same [68]. For this reason, we calculate the recoil velocity from the full waveform by using the momentum flux integral. For the final mass estimate, we use the value indicated at the metadata file.

---

[3] In particular, we use `SimPhenomUtilsIMRPhenomDFinalMass` [55, 56] for PhenomHM, `SimIMRPhenomXFinalMass2017` [20] for PhenomXHM, `SimIMREOBFinalMassSpinPrec` [65] for SEOB-NRv4HM_ROM and `NRSur3dq8Remnant` [44] for NRHybSur3dq8.

Finally, we also employ the NR Sur fit [44], which estimates the final properties of the remnant black hole $\{m_f, \vec{x}_f, \vec{v}_f\}$ from the initial intrinsic properties of the binary $\{m_1, m_2, \vec{\chi}_1, \vec{\chi}_2\}$. The surrogate model is trained on quasicircular NR simulations using Gaussian process regression. The NR Sur fit includes two surrogate models: the NRSur7dq4Remnant fit, trained on precessing systems with $q \leqslant 4$ and $|\vec{\chi}_1| = 0.8$, $|\vec{\chi}_2| = 0.8$, and the NRSur3dq8Remnant fit, trained on nonprecessing systems with $q \leqslant 8$ and $|\vec{\chi}_1| = 0.8$, $|\vec{\chi}_2| = 0.8$.

The kick prediction is in general limited by two aspects. First, waveform models only include a finite number of harmonics. Phenom models include a subset of the spherical harmonics with $\ell \leqslant 4$, while SEOBNRv4HM_ROM and the NRHybSur3dq8 include a subset of the harmonics up to $\ell \leqslant 5$ (see table 1). What we use as the gravitational strain is given by:

$$\lim_{r \to \infty} rh = \sum_{\ell=2}^{\ell_{max}} \sum_{m=-\ell}^{\ell} h_{\ell,m}(t,\lambda) \; {}^{-2}Y_{\ell,m}(\theta,\varphi). \tag{8}$$

However, the radiated momentum flux is related to the sum over an infinite number of subdominant harmonics. The omission of specific harmonics can influence the final kick velocity value.

Second, since waveforms are only available for a limited time range, the time integration of equation (5) is truncated to a finite time. Because the GW amplitude decreases exponentially in the ringdown, having a finite upper limit does not influence the final velocity value. The lower bound limits the amount of early inspiral phase included in the waveform. Based on PN calculations, there exist expressions of the net linear momentum radiated during the early inspiral phase [70–72]. The contribution of the inspiral phase to the total linear momentum is significantly smaller than the merger contribution. For a non-spinning binary with symmetric mass ratio 0.2 and total mass $50 M_{\odot}$, the radiated linear momentum up to $10\,\mathrm{Hz}$ is less than $0.05\,\mathrm{km\,s^{-1}}$. In our calculations we neglect the linear momentum radiated up to $10\,\mathrm{Hz}$.

We apply the methods discussed to calculate the estimates of the kick velocity from the previously specified waveform models over the parameter space. We divide our results into three parts. We first make model-model comparisons of the magnitude and orientation of the kick velocity. Then, we quantify where exactly in the parameter space models show disagreements by analysing the dependence of the kick estimates on the mass ratio and the individual spin components. We then compare the harmonic contributions of the kick velocity estimated by the models.

Even though NR simulations provide the closest description of the true waveform, only a limited set of NR waveforms is available. In the region where binaries have highly asymmetric masses, simulations are particularly sparse. Therefore, studying the agreement with respect to a particular model allows us to make a more exhaustive analysis of the differences between model estimates over the parameter space. For this reason, we have analysed model-model agreement by choosing a reference model and calculating the relative difference with respect to its estimates.

Because the NRHybSur3dq8 model includes a larger set of subdominant harmonics, we are able to compare all the harmonic contributions predicted by the models (each model has a particular set of harmonic contributions, different from the other models) to the NR Sur model estimates. The NR Sur is slightly more accurate than the Phenom and SEOBNR models. These models, on the other hand, cover a wider range of the parameter space than the NR Sur, relevant for LVK data analysis purposes.

However, since the true waveform is unknown, the actual value of the kick velocity is also unknown. By comparing the predictions of two different models, it is difficult to tell which of the two estimates is more accurate. However, disagreements between models are a reflection of systematic errors in either or both waveform models.

The direction of the kick velocity is subject to the orientation of the binary at the initial reference frame. In order to compare the estimates of the direction of the kick velocity from different models, we need to make sure that all waveforms are initially aligned in phase the same way. We first align the waveforms in time, setting $t = 0$ at the maximum amplitude of the $(2,2)$ spherical harmonic. We then apply a phase shift to each $(\ell, m)$ harmonic equal to $\Delta\phi_{\ell,m} = m/2 \times \phi_{2,2}$, where $\phi_{2,2}$ is the optimal phase shift of the $(2,2)$ spherical harmonic, obtained from the match calculation between the model of interest and the reference model. The match is expressed in terms of the Wienner inner product, defined as:

$$\langle h_1, h_2 \rangle = 4 \Re \int_{f_{\min}}^{f_{\max}} \frac{\tilde{h}_1(f)\, \tilde{h}_2^*(f)}{S_n(f)} \, \mathrm{d}f, \tag{9}$$

where $\tilde{h}$ indicates the Fourier transform of $h$ and $S_n(f)$ is the one-sided power spectral density of a GW detector. When aligning two waveforms in phase, we consider a flat noise sensitivity $S_n = 1$. The match $\mathcal{M}$ is defined as the normalized inner product maximized over relative time and phase shifts between the two waveforms, i.e.:

$$\mathcal{M}(h_1, h_2) = \max_{t_0, \varphi_0} \frac{\langle h_1, h_2 \rangle}{||h_1||\, ||h_2||}. \tag{10}$$

The mismatch $\mathcal{MM}$ is then defined as:

$$\mathcal{MM}(h_1, h_2) = 1 - \mathcal{M}(h_1, h_2). \tag{11}$$

## 3. Results

### 3.1. Model-model comparisons

We calculate the estimates of the kick velocity from the indicated waveform models for discrete points in the parameter space. The spin components are uniformly distributed by selecting points in the interval $\chi_{1,2}^z = [-0.8, 0.8]$ with step size $\Delta\chi_{1,2}^z = 0.1$. We consider masses $m_1$ and $m_2$ subject to the symmetric mass ratio and total mass values. Because the estimate of the kick velocity does not depend on the total mass, we only sample over the symmetric mass ratio. The symmetric mass ratio is sampled uniformly by choosing 31 points in the interval $\eta = [0.10, 0.25]$ with step size $\Delta\eta = 0.005$, giving a total number of 8959 points in the parameter space. The region of the parameter space that we choose to analyse is limited by the region to which the NRHybSur3dq8 model is calibrated.

In figure 1 we show the differences of the kick magnitude predicted by the PhenomHM, PhenomXHM and SEOBNRv4HM_ROM waveform models compared to the NRHybSur3dq8 model. We observe that the differences of PhenomXHM and SEOBNRv4HM_ROM have comparable values, with SEOBNRv4HM_ROM showing better agreement with the NRHyb-Sur3dq8 model. The distribution of PhenomXHM has a mean value of $\Delta v \sim 9\,\mathrm{km\,s^{-1}}$ and a standard deviation (std) of $\sigma \sim 19\,\mathrm{km\,s^{-1}}$. For SEOBNRv4HM_ROM, the mean value lies at $\Delta v \sim -0.4\,\mathrm{km\,s^{-1}}$ and the std $\sigma \sim 19\,\mathrm{km\,s^{-1}}$. On the other hand, the PhenomHM model largely over- and underestimates the kick velocity, with a mean value of $\Delta v \sim 42\,\mathrm{km\,s^{-1}}$ and a standard deviation of $\sigma \sim 81\,\mathrm{km\,s^{-1}}$. Figure 1 shows that PhenomHM has been superseded in accuracy by its respective newer version, PhenomXHM.

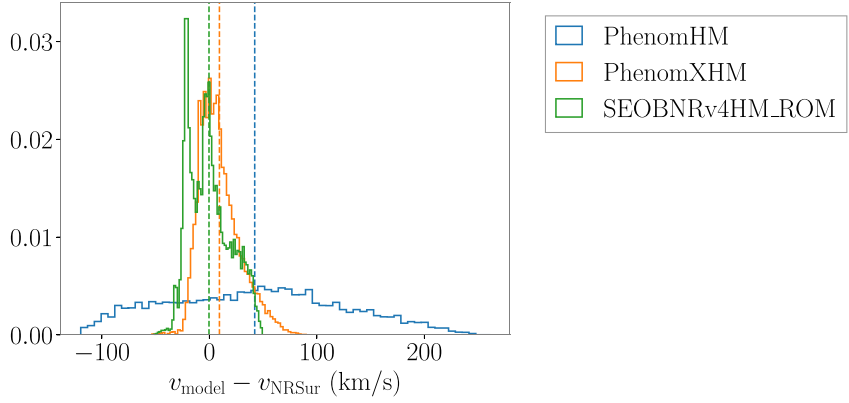

**Figure 1.** Distribution of the differences on the estimates of the kick magnitude between the waveform models indicated at the panel and the NRHybSur3dq8 model. We have considered nonprecessing black-hole binary configurations with $0.1 < \eta < 0.25$ and $-0.8 < \chi_{1,2}^z < 0.8$. Dashed vertical lines indicate the mean value of each normalised distribution.

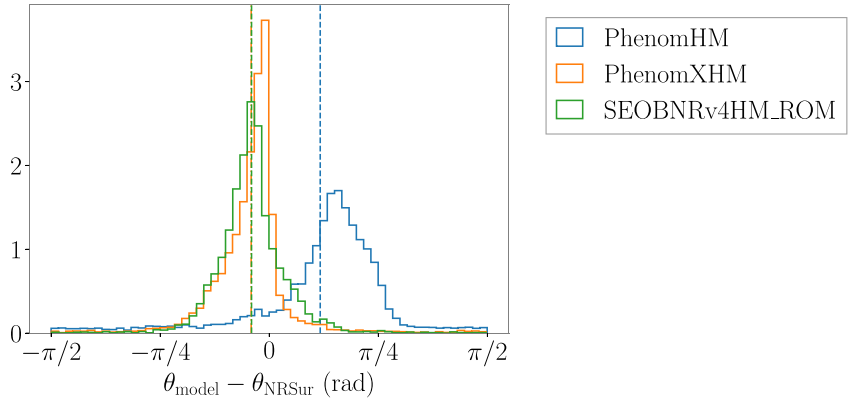

**Figure 2.** Distribution of the differences on the estimates of the kick orientation between the waveform models indicated at the panel and the NRHybSur3dq8 model. We have considered nonprecessing black-hole binary configurations with $0.1 < \eta < 0.25$ and $-0.8 < \chi_{1,2}^z < 0.8$. Dashed vertical lines indicate the mean value of each normalised distribution.

We now discuss the distributions of the differences on the direction of the kick velocity, shown in figure 2. In this case we observe good agreement between PhenomXHM and SEOB-NRv4HM_ROM with the NRHybSur3dq8 model in the kick orientation, with the mean value at $\Delta \theta \sim -0.13\,\mathrm{rad}$ and $\Delta \theta \sim -0.13\,\mathrm{rad}$ respectively. For PhenomHM, the distribution appears shifted by $\Delta \theta \sim 0.4\,\mathrm{rad}$, meaning that the orientation estimates are inconsistent with the three other waveform models.

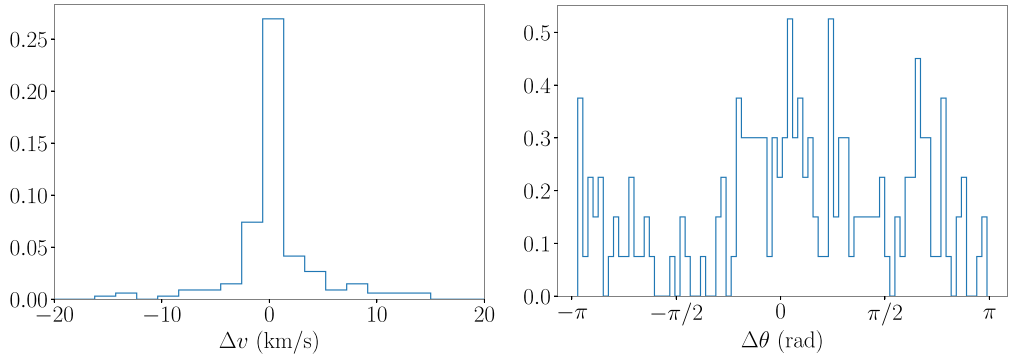

**Figure 3.** Error estimate of the kick magnitude (left) and orientation (right) for 173 nonprecessing SXS waveforms coming from the NR resolution error.

The NRHybSur3dq8 model is trained against 104 hybridized nonprecessing NR waveforms [61] and inherits the accuracy limitations of the hybrid waveforms. We now investigate the uncertainty of the kick estimate in NR simulations. These waveforms contain two primary sources of error: numerical resolution and extrapolation errors. In the case of SXS waveforms, the uncertainty involved in the extrapolation procedure is, on average, significantly smaller than the resolution error [68]. For this reason, here we focus on understanding the influence of the resolution error in our kick estimates. We estimate the kick uncertainty by comparing the kick values of the highest two resolutions. In particular, we have used 173 nonprecessing NR waveforms from the SXS Collaboration, which include at least two resolutions. The set of waveforms we employ cover a larger region of the parameter space than the set used to calibrate the NRHybSur3dq8, with $q \leqslant 10$ and $\chi^2_{1,2} \leqslant 0.994$. We provide a list of the waveforms employed in appendix B. The error estimates of the kick magnitude and orientation are shown in figure 3.

The distribution of the kick-magnitude uncertainty has a std of $\sim 10 \mathrm{~km\,s^{-1}}$. While the errors of waveform models have values between $(-\pi/2, \pi/2)$, the orientation errors of the NR waveforms lie between $(-\pi, \pi)$. The maximum value is set by our methodology and is different in each case. In the case of waveform models, we align the spherical harmonics with a phase shift $\Delta\phi_{\ell,m} = m/2 \times \phi_{2,2}$. Here, the phase of the $(2,2)$ harmonic is degenerate:

$$\phi_{2,2} = \phi^{opt}_{2,2} + 2\pi n, n \in \mathbb{Z}, \tag{12}$$

where $\phi^{opt}_{2,2}$ is the optimal phase shift of the $(2,2)$ spherical harmonic. This means that the phase shift is also degenerate:

$$\Delta\phi_{\ell,m} = m/2 \times \phi^{opt}_{2,2} + \pi n, n \in \mathbb{Z}. \tag{13}$$

Applying a phase shift $\Delta\varphi$ to a waveform translates as rotating the kick orientation by $\Delta\varphi$. Therefore, when applying $\Delta\phi_{\ell,m}$ to each spherical harmonic, the kick orientation acquires a $\pi n$ degeneracy. For this reason, the orientation estimates of the waveform models take values in the range $\theta \in (-\pi/2, \pi/2)$ and thus, the orientation differences also lie between $(-\pi/2, \pi/2)$. With NR waveforms, however, there is no such degeneracy and the orientation differences lie between $(-\pi, \pi)$.

What is important here is that both waveform models and NR simulations have orientation errors that extend to the maximum values. In the case of NR waveforms, we measure a standard deviation of ∼0.8 rad. The large uncertainty of the kick orientation is mostly related to the fact that, for each waveform resolution, the dominant harmonic contribution, which comes from the $(2, \pm 1)(2, \pm 2)$ pair of harmonics, has a significantly different kick orientation in each case.

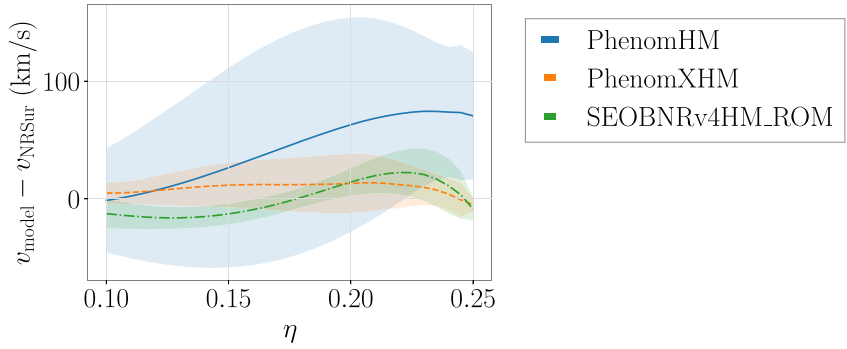

**Figure 4.** Comparison of the differences on the estimates of the kick magnitude between the models indicated at the panel and the NRHybSur3dq8 model. The shaded region represents the std of such distribution at each symmetric-mass-ratio value, while the curves represent the mean value of the distributions.

### 3.2. Symmetric-mass-ratio dependency

We now proceed to find the regions in parameter space where the models show larger disagreement, reflecting the existence of waveform modelling inaccuracies in one or both of the models. We use the same data as in section 3.1 and study the dependency on the symmetric mass ratio. Figure 4 shows the differences on the magnitude of the kick velocity as a function of the symmetric mass ratio, while figure 5 shows the differences on the kick orientation. The curves represent the mean value of the distributions at each symmetric-mass-ratio value, while the shaded region represents the std in each case.

Similar to figure 1, in figure 4 we observe that the distributions of PhenomXHM and SEOB-NRv4HM_ROM have comparable values for the kick magnitude. They both show the larger differences in the region between $\eta \in [0.20, 0.25]$. We observe that the differences between PhenomHM and NRHybSur3dq8 are significantly larger than for PhenomXHM and SEOB-NRv4HM_ROM. In particular, the estimates of PhenomHM deteriorate in accuracy with increasing symmetric-mass-ratio values, showing the largest inconsistencies close to the equal-mass case.

Regarding the estimates of the kick orientation, in figure 5 we observe a constant close-to-zero mean and std for SEOBNRv4HM_ROM. Although the estimates of PhenomXHM are comparable to those of SEOBNRv4HM_ROM, they show a slightly more complicated correlation with the symmetric mass ratio, with small changes in the width of the distributions. Similar to the figure 2, the estimates of PhenomHM appear to be shifted by ∼$\pi/8$ rad in the region $\eta \in [0.10, 0.20]$, and show the largest std from the NRHybSur3dq8 values. The estimates converge to zero towards the equal-mass case.

In appendix A, we show the harmonic contributions of the kick velocity as a function of the symmetric mass ratio, which help to understand the origin of the broad distributions observed for PhenomHM. We also include the symmetric-mass-ratio dependency of the NR errors in appendix C.

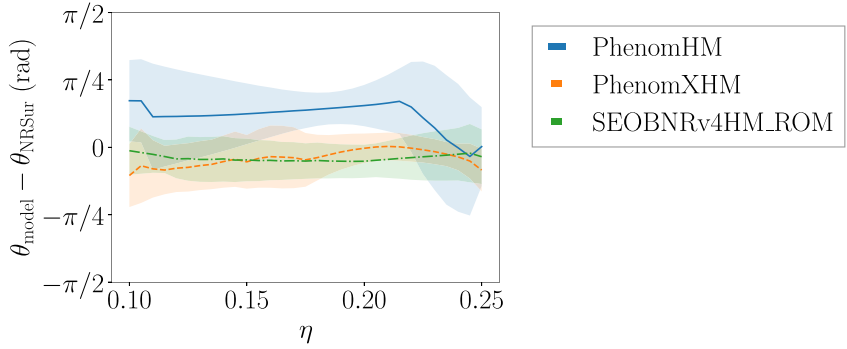

**Figure 5.** Comparison of the differences on the estimates of the kick orientation between the models indicated at the panel and the NRHybSur3dq8 model. The shaded region represents the std of such distribution at each symmetric-mass-ratio value, while the curves represent the mean value of the distributions.

In figure 6, we show the kick magnitude as a function of the symmetric mass ratio for two fixed spin configurations: nonspinning (left column) and positively highly spinning (right column). We analyse the kick estimates of the four waveform models for such configurations. In addition, we include the estimates of the two NR Sur fits, namely NRSur7dq4Remnant and NRSur3dq8Remnant and the estimates of a set of SXS waveforms.

In the case of nonspinning binaries (left column), we observe good agreement between the SXS waveforms and the NR Sur models. The estimates of the Phenom and SEOB-NRv4HM_ROM models show disagreement in the region $\eta \in [0.15, 0.24]$. The highly spinning configurations (right column) show larger relative errors between models. In particular, we observe a secondary maximum in the estimates of PhenomXHM and SEOBNRv4HM_ROM, located in the region $\eta \in [0.05, 0.15]$. Even if the true values are not known, it is highly probable that PhenomHM largely overestimates the value of the kick velocity, since the higher harmonics that induce the kick are not calibrated to NR in this model.

We now illustrate how sensitive the kick estimate is compared to the mismatch uncertainty between the models and the NR waveforms. We choose the BBH configuration $\{\eta = 0.22, \chi_1^z = 0.0, \chi_2^z = 0.0\}$. Such binary lies within the region where the models studied here have been calibrated to NR simulations. For this reason, we expect the mismatch errors with respect to the NR waveform to be small. We have calculated the mismatch for the plus ($h_+$) and cross ($h_\times$) polarizations between the waveform models and the SXS waveform, SXS:BBH:0169, for three inclination values: 0, $\pi/3$ and $\pi/2$ (rad). We have considered the Advanced-LIGO design sensitivity curve [73] with a lower cutoff of $f_{\min} = 10$ Hz. When computing the mismatch we maximise the overlap over the relative phase. Here we do not optimise this quantity over the relative phase, since these waveforms include higher harmonics and for such waveforms, a relative phase shift does not leave the waveform invariant.

Table 2 displays the mismatch values and kick differences of the models with the SXS waveform. From the mismatch values we are tempted to conclude that the waveform models are highly accurate for such particular binary configuration. However, the large disagreement in the kick estimates indicates the existence of modelling errors in the description of the GW harmonics during the merger phase. These results reflect the sensitivity of the kick velocity to waveform systematic errors.

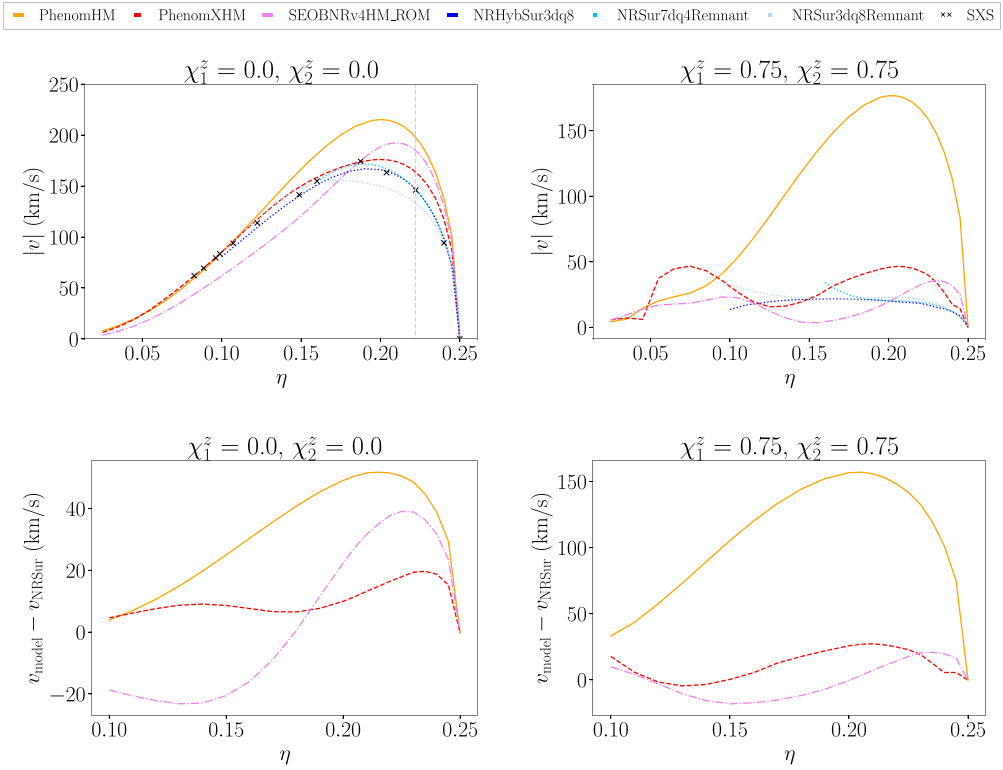

**Figure 6.** The top figures show the magnitude of the kick velocity as a function of the symmetric mass ratio estimated by PhenomHM (orange), PhenomXHM (red), SEOBNRv4HM_ROM (pink), NRHybSur3dq8 (dark blue), NRSur7dq4Remnant (skyblue) and NRSur3dq8Remnant (light blue), for two specific binary configurations. At each panel, the individual dimensionless spin components have a fixed value, specified on the top of each plot. The bottom panels show the difference of the kick magnitude between the indicated waveform model and the NRHybSur3dq8 model, for the same binary configurations of the corresponding top panels.

**Table 2.** Mismatch values and kick differences of the waveform models with SXS:BBH:0169, which has the intrinsic properties $\{\eta = 0.22, \chi_1^z = 0.0, \chi_2^z = 0.0\}$. We have computed mismatches for three different inclination angles. Here we show the minimum and maximum mismatch values obtained.

| Waveform model | $\mathcal{MM}$ | $\Delta v \, (\mathrm{km\,s^{-1}})$ |
|---|---|---|
| SEOBNRv4HM_ROM | $(0.002, 0.047)$ | 42 |
| PhenomXHM | $(0.005, 0.008)$ | 20 |
| PhenomHM | $(0.008, 0.045)$ | 55 |
| NRHybSur3dq8 | $(0.007, 0.011)$ | 4 |
| NRSur3dq8Remnant | — | 10 |
| NRSur7dq4Remnant | — | 4 |

### 3.3. Spin dependency

After analysing the mass ratio dependency, we now study whether there is any correlation between the waveform predictions and the spin components of the binary. Using the same data, we compute the differences with the NRHybSur3dq8 model as a functions of the individual spins and calculate the distributions' mean value. Our results are shown in figure 7. The left column shows the differences of the kick magnitude for PhenomHM (first row), PhenomXHM (second row) and SEOBNRv4HM (third row) with NRHyb-Sur3dq8. The right column displays the differences of the estimates on the kick orientation for the same models. In addition, we include the spin dependency of the NR errors in appendix C.

We observe discrepancies between all models in the magnitude, direction and overall dependency concerning the spin components. Regarding the differences with the NRHyb-Sur3dq8 model on the kick magnitude (left column), in the case of PhenomHM, we observe a correlation with the spin component of the more massive object. The larger the absolute magnitude of the primary spin component is, the larger the difference appears to become. PhenomXHM shows larger disagreement with the NRHybSur3dq8 model in the regions where both individual objects have the largest positive and negative spin magnitudes. In particular, we observe significantly larger relative errors in the region where both objects have negative spin components. As expected, we observe that the newer model is superior to the previous Phenom model. We observe that from the three models, it is SEOBNRv4HM_ROM the model with the closest predictions to those of NRHybSur3dq8. The largest disagreement between SEOBNRv4HM_ROM and NRHybSur3dq8 appear in the region where both spin components have large positive values and where the spin component of the more massive object has the largest negative values.

The right column in figure 7 displays the differences of the orientation estimates. In the case of the PhenomHM model, we observe discrepancies larger than $\pi/8$ rad with NRHybSur3dq8 for binaries with primary spin values of $\chi_1^z \in [-0.8, 0.4]$. For configurations with primary spin values of $\chi_1^z \in [0.4, 0.8]$, it shows good agreement with NRHybSur3dq8. PhenomXHM and SEOBNRv4HM_ROM show similar correlation with the primary spin component. The larger the magnitude of the primary spin component is, the larger inconsistencies we observe. This is more strongly reflected for the SEOBNRv4HM_ROM model, where those binaries with $\chi_1^z \in [0.4, 0.8]$ and $\chi_2^z \in [-0.8, 0.0]$ show a disagreement of $\Delta\theta \sim \pi/16$ rad with the NRHyb-Sur3dq8 model. In the case of the PhenomXHM, such correlation appears to be more subtle. The regions where the differences are around $\Delta\theta \sim \pi/16$ rad are significantly smaller, with an outlier at $\{\chi_1^z = 0.7, \chi_2^z = -0.3\}$.

### 3.4. Harmonic decomposition of the kick velocity

The kick velocity results from the sum of the contributions coming from pairs of GW harmonics. We decompose the kick velocity into its harmonic contributions, which allows us to look in more detail into the GW spherical harmonics included in the waveform models. Looking at the individual harmonic contributions allows us to understand which harmonics show more significant disagreements and, in turn, indicate the presence of waveform systematics on a more detailed level.

Equation (5) relates two GW harmonics with different $(\ell, m)$ numbers. Besides, the contributions of the pairs $(\ell_1, m_1)(\ell_2, m_2)$ and $(\ell_2, -m_2)(\ell_1, -m_1)$ have the same magnitude and direction. The dominant contribution always comes from the $(2,1)(2,2)$ and $(2,-2)(2,-1)$ pairs, and we refer to them jointly as $(2, \pm 1)(2, \pm 2)$. Besides, the number of GW harmonic contributions to the kick velocity will vary from one waveform model to another, depending on the number of spherical harmonics included in each model. We have calculated the harmonic contributions individually and compared the estimates to those of the NRHybSur3dq8 model for the same set of points as in the previous section, which are uniformly sampled in the parameter space.

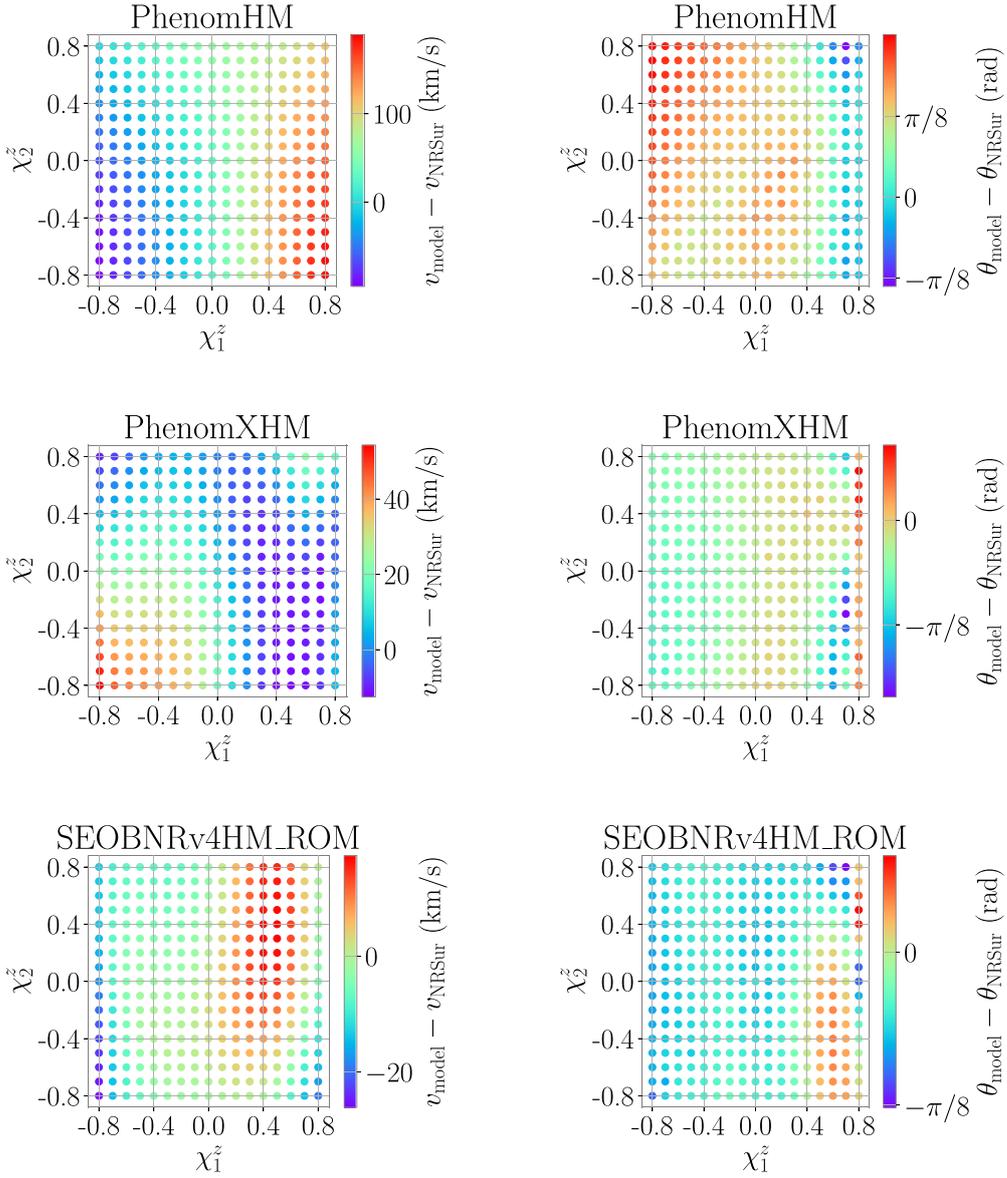

**Figure 7.** Comparison of the differences on the kick magnitude (left column) and orientation (right column) between PhenomHM (first row), PhenomXHM (second row), SEOBNRv4HM_ROM (third row) and the NRHybSur3dq8 model. Here we plot our results in terms of the individual spins. Each point of the plot is the mean value of a distribution of kick differences, where the spins have a fixed value and the mass ratio has values which have been sampled uniformly. We note that each panel has a different colormap scale.

Figure 8 shows the differences in the kick magnitude of four harmonic contributions. The pairs of spherical harmonics are indicated at the top of each plot. SEOBNRv4HM_ROM and PhenomXHM show comparable differences to the NRHybSur3dq8 estimates, while PhenomHM largely under- and overestimates the magnitude of the kick velocity. The largest differences appear for contributions where an $(\ell, \pm(\ell - 1))$ harmonic is involved. PhenomXHM is in good agreement with NRHybSur3dq8, particularly for the $(3, \pm2)(3, \pm3)$ and $(3, \pm3)(4, \pm4)$ contributions. We should note that the SEOBNRv4HM_ROM model does not include the $(3, 2)$ harmonic, and thus, it has no contribution coming from the $(3, \pm2)(3, \pm3)$ pair. In the case of SEOBNRv4HM_ROM, we observe good agreement with the NRHyb-Sur3dq8, particularly for the $(2, \pm1)(2, \pm2)$ and $(2, \pm2)(3, \pm3)$ contributions.

Figure 9 displays the distribution of the differences in the orientation of the kick velocity. Here we observe a similar pattern to the one displayed in figure 2. The distributions of the differences are centered around zero for SEOBNRv4HM_ROM with $\sigma \sim 0.3$ rad, showing good agreement with the NRHybSur3dq8 model. In the case of the PhenomXHM estimates, we observe good agreement with NRHybSur3dq8, with $\sigma \sim 0.4$ rad, except for the $(3, \pm2)(3, \pm3)$ harmonic contribution, where the distribution has a significantly larger std, $\sigma \sim 0.6$ rad. The distributions of PhenomHM appear shifted at every contribution, with $\sigma \sim 0.8$ rad. In particular, the $(3, \pm2)(3, \pm3)$ and $(3, \pm3)(4, \pm4)$ contributions, show a significantly larger disagreement with the NRHybSur3dq8 model.

## 4. Towards kick-based GW tuning

Because the kick estimate is a highly sensitive quantity, here we investigate whether we can use the kick prediction of NR simulations to add additional information in the calibration of the EOBNR and Phenom waveform models. These models calibrate unknown coefficients to NR simulations for the merger and ringdown. Unlike the inspiral region, where the amplitude and phase of the signal are well known from PN or EOB theory, modelling inaccuracies might build up in the merger region. In the following, we propose the first steps towards incorporating the kick prediction in the calibration of a waveform model.

Ideally, one could try to improve the description of the radiated linear momentum together with the energy and angular momentum by tuning these three quantities together. However, this would significantly increase the complexity of the problem. The energy and angular momentum are dominated by the $(2, \pm2)$ modes, while the linear momentum is dominated by the interplay of the $(2, \pm2)$ with other higher harmonics. The EOBNR and Phenom waveform families started with models for the dominant mode, which over the years have been improved by increasing the number of NR waveforms used for the calibration. For this reason, we expect the energy and angular momentum to be better modeled than the kick velocity, which has a more subtle imprint in the waveform. As a first step, one would then focus on the kick description, making sure that the radiated energy and angular momentum are not modified dramatically. Expressions for the energy and angular momentum in terms of the radiated harmonics can be found in [66].

As indicated before, the prediction of the kick orientation in currently available NR waveforms has a large uncertainty ($\sigma \sim 0.8$ rad). For this reason, it might be difficult to significantly improve the performance of waveform models with current NR waveforms. Here we discuss how we could use information from the kick estimate once we have more accurate NR waveforms, and in turn, more precise kick estimates. By comparing the kick predictions of a waveform model with those predicted by NR, we analyse whether it is possible to carefully tune the individual harmonic contributions and, in turn, improve the accuracy of the modelled gravitational signal. We discuss the requirements that must be fulfilled and describe the complexity of addressing such a problem.

When analysing GW data, small phase variations are much better measured compared to amplitude variations. Hence, most of the information that is inferred from the waveform comes from the phase. The first question we want to answer is, whether it is possible to calibrate the GW phase based on the kick contributions.

We tune the harmonics contributions applying a phase shift $\alpha_{\ell,m}$ to the GW harmonic $h_{\ell,m}$, such that $h_{\ell,m} = A_{\ell,m}(t)\,e^{-i\phi_{\ell,m}(t)}$ transforms into:

$$\bar{h}_{\ell,m} = A_{\ell,m}(t)\,e^{-i\bar{\phi}_{\ell,m}(t)}, \tag{14}$$

where $\bar{\phi}_{\ell,m}(t) = \phi_{\ell,m}(t) + \alpha_{\ell,m}(t,\lambda)$ is the tuned phase function. As specified, in principle, $\alpha_{\ell,m}(t,\lambda)$ can be a function of the time evolution and the intrinsic parameters of the binary. We can also write the waveform simply as:

$$\bar{h}_{\ell,m} = h_{\ell,m}(t)\,e^{-i\alpha_{\ell,m}(t,\lambda)}. \tag{15}$$

In reality, we are interested in the kick prediction, which involves the first time derivative of the gravitational strain:

$$\dot{h}_{\ell,m} = \dot{A}_{\ell,m}(t)\,e^{-i\phi_{\ell,m}(t)} - i\,h_{\ell,m}\dot{\phi}_{\ell,m}(t). \tag{16}$$

The time derivative of the transformed waveform is on the other hand,

$$\dot{\bar{h}}_{\ell,m} = \left(\dot{h}_{\ell,m}(t) - i\,h_{\ell,m}(t)\dot{\alpha}_{\ell,m}(t,\lambda)\right)e^{-i\alpha_{\ell,m}(t,\lambda)}. \tag{17}$$

### 4.1. Applying a constant phase shift

In the simple case where the applied phase shift is slowly varying, such that its time derivative can be neglected, the first derivative of the spherical harmonic reduces to:

$$\dot{\bar{h}}_{\ell,m} = \dot{h}_{\ell,m}(t)\,e^{-i\alpha_{\ell,m}}. \tag{18}$$

We now look at the the kick formula and compute the kick contribution of the pair of harmonics $(\ell,m)$ and $(\ell,m+1)$, each with a constant phase shift $\alpha_{l,m}$ and $\alpha_{l,m+1}$ respectively. We have:

$$\bar{v}_{(\ell,m)\,(\ell,m+1)} = -\frac{1}{8\pi M_f}\int_{-\infty}^{\infty}\mathrm{d}t\left(\dot{h}_{\ell,m}(t)\,e^{-i\alpha_{\ell,m}}\right)a_{\ell,m}\left(\dot{h}_{\ell,m+1}(t)\,e^{-i\alpha_{\ell,m+1}}\right)^{*}. \tag{19}$$

Since the phase shift is simply a constant, the exponential term can be taken out of the integral. We can see that the transformed kick contribution is rotated by the difference of the constant phase shifts:

$$\bar{v}_{(\ell,m)\,(\ell,m+1)} = v_{(\ell,m)\,(\ell,m+1)}\,e^{-i(\alpha_{\ell,m}-\alpha_{\ell,m+1})}. \tag{20}$$

This means, a constant phase shift allows us to perfectly correct the orientation of the contributions, and keep the amplitude unchanged at the same time. In general, we consider two $(\ell_1,m_1)$ and $(\ell_2,m_2)$ harmonics, with a phase shift $\alpha_{\ell_1,m_1}$ and $\alpha_{\ell_2,m_2}$, respectively. Their kick contribution will be shifted by:

$$\Delta\theta_{\mathrm{kick}} = \alpha_{\ell_2,m_2} - \alpha_{\ell_1,m_1}. \tag{21}$$

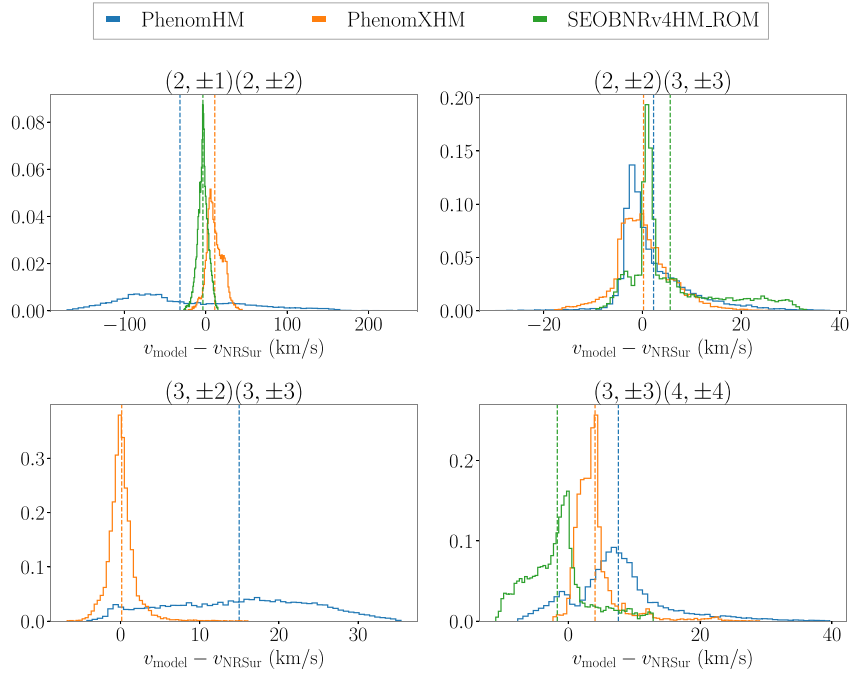

**Figure 8.** Harmonic contributions of the kick magnitude. We show results for PhenomHM (blue), PhenomXHM (orange) and SEOBNRv4HM_ROM (green) compared to NRHybSur3dq8. The pairs of GW harmonics are specified at the top of each panel. Dashed vertical lines indicate the mean value of each normalised distribution.

Now, the contributions from the two pairs $(\ell_1, m_1)(\ell_2, m_2)$ and $(\ell_2, -m_2)(\ell_1, -m_1)$ have the same magnitude and orientation. So, in the same way, we have:

$$\Delta\theta_{\text{kick}} = \alpha_{\ell_1, -m_1} - \alpha_{\ell_2, -m_2}. \tag{22}$$

Therefore, if the calibration phase shift of a particular harmonic $\alpha_{\ell, m}$ is known, the phase shift of the $(\ell, -m)$ harmonic will be directly determined. Thus, we only need to find the required phase shift for half of the harmonics included in the model. In the following, we use PhenomHM as the base model that we want to recalibrate. However, the procedure described next can be generalised to any waveform model. The reason why we choose PhenomHM is simply that accuracy improvements might be easier to appreciate. Looking at figures 8 and 9, we observe that the $m = \ell - 1$ harmonics, namely the $(2, 1)$, $(3, 2)$ and $(4, 3)$, are the least accurately modelled for PhenomHM. These are the harmonics we might want to tune.

In the case of PhenomHM, there are 7 kick contributions that we want to correct. The model includes 6 (positive $m$) harmonics. In addition, there is one extra degree of freedom that is the reference orbital phase, which is fixed in each SXS waveforms, but not for the Phenom models. The set of unknown phase shifts can be determined by solving a linear system of equations for the positive (or negative) $m$ harmonics.

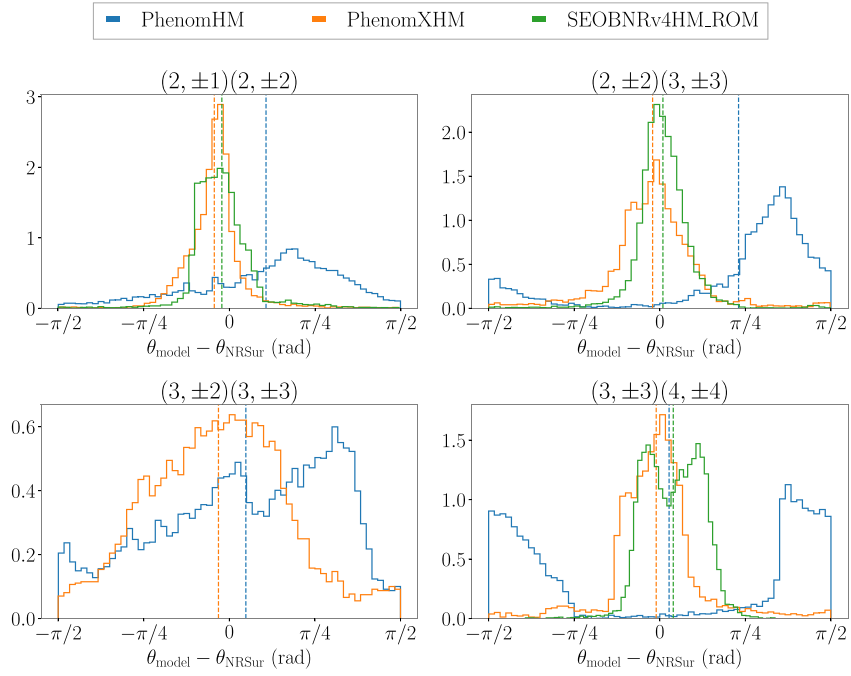

**Figure 9.** Harmonic contributions of the kick orientation. We show results for PhenomHM (blue), PhenomXHM (orange) and SEOBNRv4HM_ROM (green) compared to NRHybSur3dq8. The pairs of GW harmonics are specified at the top of each panel. Dashed vertical lines indicate the mean value of each normalised distribution.

The linear system of equations, $Ax = b$, is given by:

$$A = \begin{pmatrix} 1 & 1 & 0 & 0 & -1 & 0 & 0 \\ 1 & -1 & 1 & 0 & 0 & 0 & 0 \\ 1 & 0 & -1 & 1 & 0 & 0 & 0 \\ 1 & 0 & 1 & 0 & 0 & -1 & 0 \\ 1 & 0 & 0 & 1 & 0 & 0 & -1 \\ 1 & 0 & 0 & 0 & -1 & 1 & 0 \\ 1 & 0 & 0 & 0 & 0 & -1 & 1 \end{pmatrix}, \tag{23}$$

$$x = (\alpha_{ref}, \alpha_{(2,2)}, \alpha_{(3,3)}, \alpha_{(4,4)}, \alpha_{(2,1)}, \alpha_{(3,2)}, \alpha_{(4,3)}), \tag{24}$$

$$b = (\theta_{(2,1)\ (2,2)}, \theta_{(2,2)\ (3,3)}, \theta_{(3,3)\ (4,4)}, \theta_{(3,2)\ (3,3)}, \theta_{(4,3)\ (4,4)}, \theta_{(2,1)\ (3,2)}, \theta_{(3,2)\ (4,3)}). \tag{25}$$

$\theta_{(2,1)(2,2)}$ refers to the orientation difference between the SXS and PhenomHM (2,1)(2,2) contribution,

$$\theta_{(2,1)(2,2)} = v_{(2,1)(2,2)}^{\text{SXS}} - v_{(2,1)(2,2)}^{\text{PhHM}}. \tag{26}$$

The solution to such a system will rotate the kick contributions of the waveform model and perfectly match those predicted by the NR waveform. The next question we want to answer is whether these shifts make the waveform model more accurate. In reality, applying a constant phase shift to the waveform would change the phase of the inspiral part, which is well modelled based on PN and EOB theory. Thus, applying a constant shift would instead make our waveforms inconsistent with the PN predictions of the early inspiral. If we want to address possible systematic errors in the waveform, we must keep the original inspiral phase and correct the merger-ringdown phase with a time-dependent phase shift.

### 4.2. Applying a time-dependent phase shift around the merger

We now want to leave the inspiral phase unchanged and apply a phase shift that is non-zero during the merger. The late-ringdown phase is determined by solving equation (23) for the fully integrated kick velocity. Ideally, one should find the phase shift $\alpha_{\ell,m}(\lambda,t)$ as a function of the intrinsic parameters for each harmonic, such that the time profile of the individual kick contributions predicted by a model is calibrated to the NR estimates. Thus, we are looking for a phase correction that is zero in the inspiral part, builds up around the merger and has a constant value in the late-ringdown.

We consider a phase shift that is time-dependent, and at least once differentiable. In this case, the time derivative of the tuned waveform involves an additional term:

$$\dot{\tilde{h}}_{\ell,m} = \left( \dot{h}_{\ell,m} - i h_{\ell,m} \dot{\alpha}_{\ell,m}(t) \right) \mathrm{e}^{-i\alpha_{\ell,m}(t)}. \tag{27}$$

The tuned kick contribution of a pair of harmonics $(\ell,m)$ and $(\ell,m+1)$, each with a time-varying phase shift $\alpha_{\ell,m}$ and $\alpha_{\ell,m+1}$ respectively, is then given by:

$$\begin{aligned}
\bar{v}_{(\ell,m)\,(\ell,m+1)} = -\frac{1}{8\pi M_f} \int_{-\infty}^{\infty} \mathrm{d}t\, [\dot{h}_{\ell,m}(t)\dot{h}_{\ell,m+1}^* + i\dot{h}_{\ell,m}h_{\ell,m+1}^*\dot{\alpha}_{\ell,m+1}^*(t) - i\dot{h}_{\ell,m+1}^*h_{\ell,m}\dot{\alpha}_{\ell,m}(t) \\
+ h_{\ell,m}h_{\ell,m+1}^*\dot{\alpha}_{\ell,m}(t)\dot{\alpha}_{\ell,m+1}^*(t)]\, a_{\ell,m}\mathrm{e}^{-i(\alpha_{\ell,m}(t)-\alpha_{\ell,m+1}(t))}.
\end{aligned} \tag{28}$$

We observe that for a phase-shift function that necessarily changes its value over time, it is not possible to recover the expression (20), where the tuned contribution is precisely rotated by the difference of the individual phase shifts. Therefore, when using a time-dependent function, we can only aim to tune the kick contributions as *close* to the NR predictions as possible.

The rest of the subsection aims to study more deeply how well we can correct the harmonic contributions. We start with a simple function with the following characteristics: it is zero for the inspiral part, builds up around the merger following a Gaussian distribution, and has a constant value in the late-ringdown. The Gaussian distribution is centred at $t/M = 0$ and has a maximum value equal to the late-ringdown phase shift. We leave the width of the distribution as a free parameter and study how the tuned kick velocity (integrated over the whole evolution) depends on its value. Figure 10 displays the function we first try, with different values of the std of the Gaussian distribution. As a first step, we should mention that one could choose a different type of distribution, centre it at a different point in time, and would still come to the same conclusions.

Next, we want to address whether a more convenient width can be chosen for all harmonics. We compute the dependency of the kick on the width for three different configurations as shown in figure 11.

Because we are only trying to correct the orientation of the contributions, it might be that even though the direction of the kick is close to the NR prediction, the individual magnitudes differ significantly from the NR estimates. A better approach should consider both amplitude and phase corrections at the same time. Besides, one should study the dependency individually for each harmonic. We observe that the value of the distribution width might change for different binary configurations. Thus, the width should not only vary for each harmonic, but one should also find dependency on the intrinsic parameters of the binary.

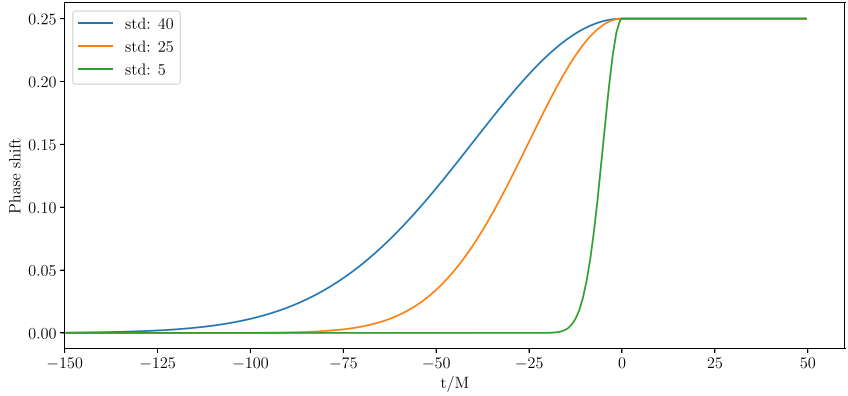

**Figure 10.** We consider a time-dependent phase shift that builds up in the merger following a gaussian distribution centred at $t/M = 0$. Initially, the width of the distributions is left free and can take different values, as shown in the figure.

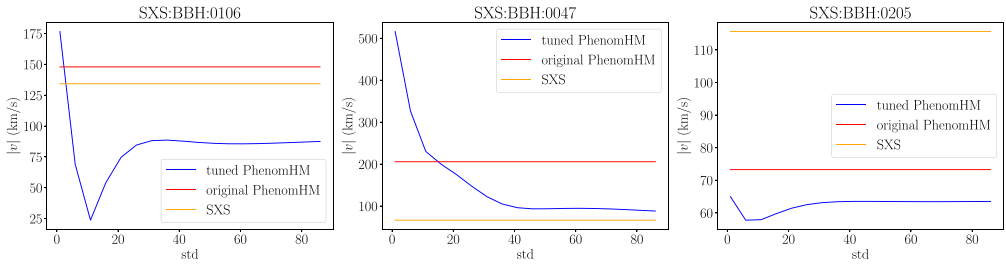

**Figure 11.** The dependency of the kick velocity on the value of the std of the Gaussian distribution is shown for three different configurations. The horizontal lines indicate the kick velocity estimate of the SXS waveform (orange) and that of the original PhenomHM waveform (red). The kick estimate of the transformed PhenomHM waveform with a phase shift with a specific standrad deviation value is shown in blue.

Although our current approach can be improved in many ways, we want to know whether our simple model can still improve in the phase of the modelled GW harmonic. We use 9 SXS waveforms of different configurations as indicated in table 3 and apply the following algorithm:

- Find the value of the std of the distribution that leads to a total kick value that matches the SXS prediction.
- Find the required late-ringdown phase shifts for each harmonic by solving the linear system of equations for the fully integrated kick contributions.
- Apply the tuning phase shift to each harmonic.

To analyse how the phase shift modifies the original PhenomHM harmonics, we plot the original and tuned waveform, its phase and the phase difference between the SXS and the PhenomHM waveforms. We include the results of one of the configurations in figures 12 and 13, where we show how the phase shifts modify the phase and in turn the waveform for the $(2,2)$ and $(4,4)$ harmonics. In all 9 studied cases, we find that the simple phase-shift function does not improve the accuracy of the modelled waveform. Just as for the constant phase shift, it is not straightforward that tuning the kick contributions necessarily implies an improvement of the waveform.

**Table 3.** SXS waveforms used in our study to calibrate the kick contributions of PhenomHM.

| Waveform ID | $\eta$ | $\tilde{\chi_1^z}$ | $\tilde{\chi_2^z}$ |
| --- | --- | --- | --- |
| SXS:BBH:0046 | 0.188 | $-0.5$ | $-0.5$ |
| SXS:BBH:0047 | 0.188 | 0.5 | 0.5 |
| SXS:BBH:0106 | 0.139 | 0.0 | 0.0 |
| SXS:BBH:0186 | 0.096 | 0.0 | 0.0 |
| SXS:BBH:0191 | 0.204 | 0.0 | 0.0 |
| SXS:BBH:0199 | 0.092 | 0.0 | 0.0 |
| SXS:BBH:0205 | 0.109 | $-0.4$ | 0.0 |
| SXS:BBH:0222 | 0.250 | $-0.3$ | 0.0 |
| SXS:BBH:0289 | 0.187 | 0.6 | 0.0 |

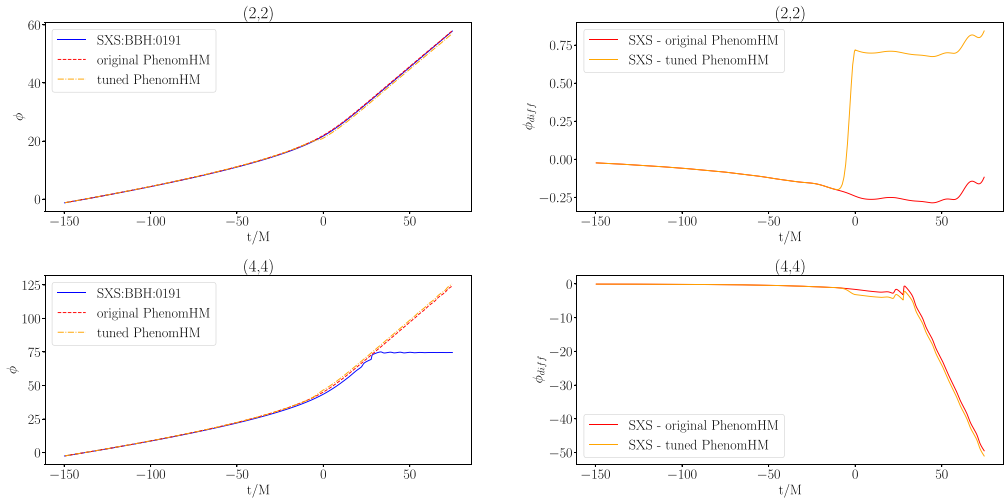

**Figure 12.** The left column displays the original and tuned PhenomHM phase compared to the SXS waveform phase, while the right column shows the phase differences between the SXS and the original (tuned) PhenomHM phase in red (orange) for the binary $\{\eta = 0.204, \chi_1 = 0.0, \chi_2 = 0.0\}$. On the top row we show the $(2,2)$ harmonic, and on the bottom the $(4,4)$ harmonic.

From the plots of the waveform phases, we find support, once again, to use a different phase function for each harmonic. We observe that the relative error in the phase does not have a simple structure which could be corrected using a phase shift that builds up following a smooth distribution. Besides, it could be that the relative errors are randomly distributed during the merger time and even over the parameter space. These two aspects reinforce our view of how complex kick-based tuning is. The main question we want to address is still how to optimally reduce the relative errors using the information contained in the kick. As mentioned earlier, the amplitude and the phase of the kick contributions should be corrected simultaneously. A time-dependent phase shift does not only change the orientation but also the magnitude of the kick contributions, as indicated by equation (28). In our research, we have only considered the change in the orientation of the kick. One could try to analyse the expression (28) in more detail and study whether it is possible to tune the amplitude and phase simply with a phase-shift function.

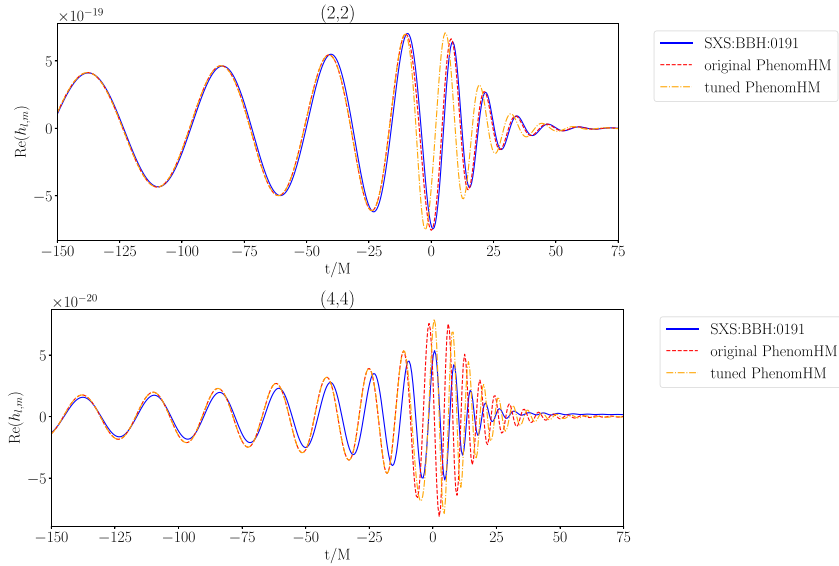

**Figure 13.** We show the $(2,2)$ and $(4,4)$ harmonics of the original (red) and tuned (orange) PhenomHM waveform together with the SXS waveform (blue) for $\{\eta = 0.204, \chi_1 = 0.0, \chi_2 = 0.0\}$.

So far, we have only calibrated the kick contributions of the late-ringdown by comparison to the fully integrated kick values. The time evolution[4] of the individual contributions could be used instead, which would allow finding the required phase shifts at several points in time, not only at the late-ringdown. One could choose a set of collocation points and solve the set of equations of the individual phases and possibly the amplitudes at each point. Besides, in our study we have explored the use of phase shifts to calibrate our estimates. However, the kick prediction is also sensitive to time shifts, which could complement the use of phase shifts. One should take care that in the waveform calibration process, the properties of the binary system are not changed. Similar to the radiated linear momentum, the energy and the angular momentum (see [37]) would have to be corrected in the models.

## 5. Discussion

In this paper, we show that current waveform models, NRHybSur3dq8, SEOB-NRv4HM_ROM, PhenomHM and PhenomXHM, are not consistent with each other in their kick estimates over the parameter space. Waveform systematic deviations that occur during the merger phase can strongly impact the kick estimate. Because the kick prediction is highly sensitive to waveform inaccuracies, disagreements between models indicate where in the parameter space are waveform systematics more significant. We have studied the dependency of the kick magnitude and orientation on the symmetric mass ratio and the individual spin components of the binary. Analysing model-model agreement, we find that overall,

---

[4] The time evolution of the kick velocity describes how the velocity of the system's center-of-mass increases with the binary evolution.

PhenomXHM and SEOBNRv4HM_ROM show comparable differences compared to NRHyb-Sur3dq8. We observe large discrepancies between the predictions of PhenomHM relative to NRHybSur3dq8. Our results show that PhenomHM has been superseded in accuracy by its newer version, PhenomXHM. Such improvement is probably related to the NR calibration of the subdominant spherical harmonics in the latest model. We observe the largest discrepancies in regions of the parameter space where the spin magnitude of the more massive black hole has high absolute values. Besides, the largest uncertainties appear in the region where the symmetric mass ratio takes values between $\eta = [0.20, 0.25]$.

Since the estimate of the kick velocity involves the description of the higher harmonics during the merger phase, we are able to study the individual contributions of the kick velocity coming from different spherical harmonics. We analyse model-model agreement and find similar results as for the total kick velocity. Both PhenomXHM and SEOBNRv4HM_ROM show considerably smaller differences to NRHybSur3dq8 than PhenomHM. In particular, we observe extremely good agreement between the PhenomXHM and NRHybSur3dq8 in the magnitude of the $(3, \pm 2)(3, \pm 3)$ and $(3, \pm 3)(4, \pm 4)$ contributions. At the same time, we find large disagreement between PhenomXHM and PhenomHM with NRHybSur3dq8 on the orientation of the $(3, \pm 2)(3, \pm 3)$ and $(3, \pm 3)(4, \pm 4)$ contribution. Our results support the use of the kick estimate as a complementary tool to the mismatch uncertainty to evaluate the performance of gravitational waveform models.

We further study whether calibrating the individual kick contributions to the NR predictions can, in turn, imply an improvement in the accuracy of a waveform model. We focus on tuning the orientations of the kick contributions, and we use PhenomHM as our base model for the study. Although applying a constant phase shift would allow us to calibrate the orientation of the individual contributions, the shift would change the well modelled inspiral phase. Instead, we need a time-dependent phase shift that builds up around the merger. However, finding the appropriate time-dependent function that corrects the waveform amplitude and phase errors is highly complex. We find that a different phase-shift function is probably needed for each harmonic, which might depend on the binary's intrinsic properties. Besides, one should treat both the magnitude and orientation of the individual kick contributions simultaneously. Once there are more accurate NR predictions of the kick orientation available, this calibration framework could help improve the modelling of the kick imprint in waveform models.

## Data availability statement

The data cannot be made publicly available upon publication because the cost of preparing, depositing and hosting the data would be prohibitive within the terms of this research project. The data that support the findings of this study are available upon reasonable request from the authors.

## Acknowledgments

The authors are grateful to Neev Khera, Abhay Ashtekar, Badri Krishnan and Cecilio García-Quirós for useful discussions. We thank Marta Colleoni for sharing the details of the kick calculations performed in [21], which we used for comparison. We thank Vijay Varma and Sascha Husa for comments on the manuscript. We are grateful to the members of the 'Binary Merger Observations and Numerical Relativity' group for useful feedback, and we thank Max Melching for sharing his code to easily display large data sets. This work was supported by the Max Planck Society's Independent Research Group program. Computations were carried out on the Holodeck cluster of the Max Planck Institute for Gravitational Physics in Hannover.

## Appendix A. Harmonic contributions as a function of the symmetric mass ratio

We further investigate our results from section 3.1 by studying the harmonic contributions of each model as a function of the symmetric-mass-ratio. Our results are displayed in figures A1 and A2, where we show the harmonic contributions of the kick magnitude and orientation respectively.

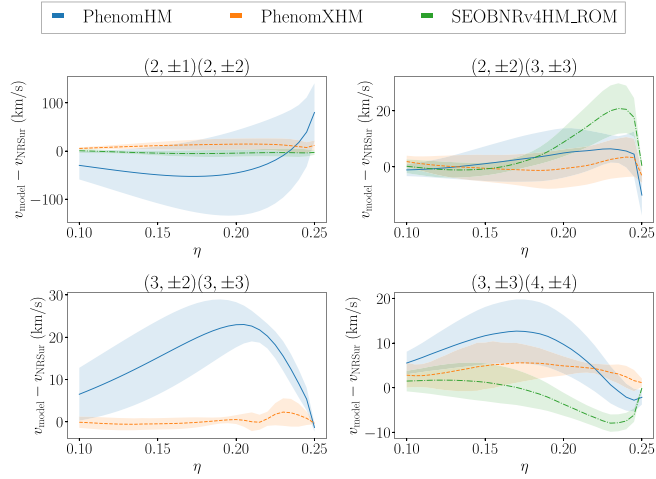

**Figure A1.** Harmonic contributions of the kick magnitude as a function of the symmetric mass ratio. We show results for the PhenomHM (blue), PhenomXHM (orange) and SEOBNRv4HM_ROM (green) models. The shaded region represents the std of such distribution at each symmetric-mass-ratio value, while the curves represent the mean value of the distributions.

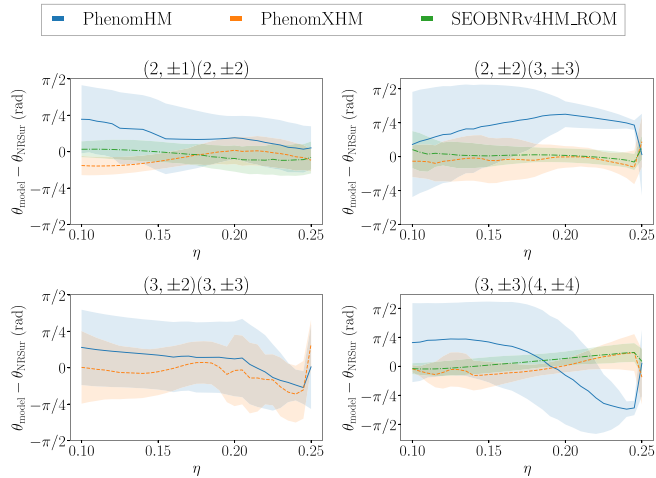

**Figure A2.** Harmonic contributions of the kick orientation as a function of the symmetric mass ratio. We show results for the PhenomHM (blue), PhenomXHM (orange) and SEOBNRv4HM_ROM (green) models. The shaded region represents the std of such distribution at each symmetric-mass-ratio value, while the curves represent the mean value of the distributions.

## Appendix B. Table of SXS waveforms employed to estimate the kick error

In table B1, we display the list of 173 nonprecessing SXS waveforms we have used to get an estimate of the kick error in NR waveforms coming from the NR resolution uncertainty. For each simulation we include the intrinsic parameters of the BBH, and the error estimate of the kick magnitude ($\Delta v$) and direction ($\Delta \theta$).

**Table B1.** Error estimates of the kick magnitude ($\Delta v$) and orientation ($\Delta \theta$) for the set of SXS waveforms indicated in the first column. For each waveform, we include the intrinsic parameters of the binary.

| Waveform ID | $\eta$ | $\chi_1$ | $\chi_2$ | $\Delta v$ | $\Delta \theta$ |
|---|---|---|---|---|---|
| SXS:BBH:0002 | 0.09 | (0.0, 0.0, 0.0) | (0.0, 0.0, 0.0) | 0.005 | −2.578 |
| SXS:BBH:0004 | 0.086 | (0.0, 0.0, 0.0) | (0.0, 0.0, 0.0) | 3.529 | −4.912 |
| SXS:BBH:0005 | 0.083 | (0.0, 0.0, 0.0) | (0.0, 0.0, 0.0) | 2.049 | 3.23 |
| SXS:BBH:0007 | 0.248 | (0.0, 0.0, 0.33) | (0.0, 0.0, −0.44) | −10.064 | 0.368 |
| SXS:BBH:0008 | 0.245 | (0.0, 0.0, 0.962) | (0.0, 0.0, −0.9) | 7.329 | 0.307 |
| SXS:BBH:0012 | 0.247 | (0.0, 0.0, 0.32) | (0.0, 0.0, −0.58) | 0.175 | 0.522 |
| SXS:BBH:0014 | 0.178 | (0.0, 0.0, 0.523) | (0.0, 0.0, −0.448) | 6.122 | −0.341 |
| SXS:BBH:0016 | 0.248 | (0.0, 0.0, 0.33) | (0.0, 0.0, −0.44) | 4.559 | 0.505 |
| SXS:BBH:0019 | 0.248 | (0.0, 0.0, 0.33) | (0.0, 0.0, −0.44) | −3.279 | −0.448 |
| SXS:BBH:0025 | 0.248 | (0.0, 0.0, 0.33) | (0.0, 0.0, −0.44) | −82.629 | 0.868 |
| SXS:BBH:0030 | 0.25 | (0.0, 0.0, 0.0) | (0.0, 0.0, 0.0) | 4.26 | −0.237 |
| SXS:BBH:0031 | 0.25 | (0.0, 0.0, −0.5) | (0.0, 0.0, 0.0) | 0.608 | 2.066 |
| SXS:BBH:0036 | 0.25 | (0.0, 0.0, 0.5) | (0.0, 0.0, 0.0) | −6.433 | 2.768 |
| SXS:BBH:0054 | 0.24 | (0.0, 0.0, 0.0) | (0.0, 0.0, 0.0) | 12.079 | 0.27 |
| SXS:BBH:0056 | 0.24 | (0.0, 0.0, 0.0) | (0.0, 0.0, 0.0) | −2.289 | −0.646 |
| SXS:BBH:0063 | 0.24 | (0.0, 0.0, −0.5) | (0.0, 0.0, 0.0) | −1.28 | −0.381 |
| SXS:BBH:0064 | 0.24 | (0.0, 0.0, −0.5) | (0.0, 0.0, 0.0) | 3.182 | 2.706 |
| SXS:BBH:0065 | 0.24 | (0.0, 0.0, −0.5) | (0.0, 0.0, 0.0) | 7.317 | 0.575 |
| SXS:BBH:0106 | 0.24 | (0.0, 0.0, −0.5) | (0.0, 0.0, 0.5) | 7.219 | 2.232 |
| SXS:BBH:0107 | 0.24 | (0.0, 0.0, 0.5) | (0.0, 0.0, −0.5) | 2.358 | −0.283 |
| SXS:BBH:0113 | 0.188 | (0.0, 0.0, 0.0) | (0.0, 0.0, 0.0) | 12.53 | 1.939 |
| SXS:BBH:0148 | 0.187 | (0.0, 0.0, 0.5) | (0.0, 0.0, 0.0) | 0.0 | −2.893 |
| SXS:BBH:0149 | 0.187 | (0.0, 0.0, −0.5) | (0.0, 0.0, 0.0) | 0.0 | 0.738 |
| SXS:BBH:0150 | 0.139 | (0.0, 0.0, 0.0) | (0.0, 0.0, 0.0) | 0.002 | 2.724 |
| SXS:BBH:0151 | 0.139 | (0.0, 0.0, 0.0) | (0.0, 0.0, 0.0) | −0.0 | −2.739 |
| SXS:BBH:0152 | 0.099 | (0.0, 0.0, 0.0) | (0.0, 0.0, 0.0) | −0.001 | 3.106 |
| SXS:BBH:0154 | 0.099 | (0.0, 0.0, −0.5) | (0.0, 0.0, 0.0) | −0.0 | 3.613 |
| SXS:BBH:0155 | 0.099 | (0.0, 0.0, 0.5) | (0.0, 0.0, 0.0) | 0.002 | 3.025 |
| SXS:BBH:0157 | 0.139 | (0.0, 0.0, 0.0) | (0.0, 0.0, 0.0) | 0.048 | −2.979 |
| SXS:BBH:0158 | 0.139 | (0.0, 0.0, 0.0) | (0.0, 0.0, 0.0) | −0.388 | −0.24 |
| SXS:BBH:0159 | 0.139 | (0.0, 0.0, 0.0) | (0.0, 0.0, 0.0) | 0.006 | 0.284 |
| SXS:BBH:0160 | 0.25 | (0.0, 0.0, −0.438) | (0.0, 0.0, −0.438) | −0.001 | 3.523 |
| SXS:BBH:0162 | 0.25 | (0.0, 0.0, −0.2) | (0.0, 0.0, −0.2) | 5.424 | −2.418 |
| SXS:BBH:0167 | 0.25 | (0.0, 0.0, 0.2) | (0.0, 0.0, 0.2) | −1.817 | 1.264 |
| SXS:BBH:0168 | 0.25 | (0.0, 0.0, −0.6) | (0.0, 0.0, −0.6) | −1.819 | −0.388 |
| SXS:BBH:0169 | 0.25 | (0.0, 0.0, 0.6) | (0.0, 0.0, 0.6) | −0.82 | −0.5 |
| SXS:BBH:0172 | 0.25 | (0.0, 0.0, −0.8) | (0.0, 0.0, −0.8) | 0.696 | −0.51 |

(Continued.)

**Table B1.** (Continued.)

| | | | | | |
|---|---|---|---|---|---|
| SXS:BBH:0174 | 0.25 | (0.0, 0.0, 0.8) | (0.0, 0.0, 0.8) | 73.582 | 2.475 |
| SXS:BBH:0175 | 0.25 | (0.0, 0.0, 0.95) | (0.0, 0.0, 0.95) | 0.205 | 3.437 |
| SXS:BBH:0176 | 0.25 | (0.0, 0.0, 0.97) | (0.0, 0.0, 0.97) | −0.231 | 2.029 |
| SXS:BBH:0177 | 0.25 | (0.0, 0.0, −0.9) | (0.0, 0.0, −0.9) | −0.223 | −0.693 |
| SXS:BBH:0178 | 0.25 | (0.0, 0.0, 0.9) | (0.0, 0.0, 0.9) | 0.002 | −0.292 |
| SXS:BBH:0180 | 0.222 | (0.0, 0.0, 0.6) | (0.0, 0.0, 0.0) | −0.003 | −1.383 |
| SXS:BBH:0181 | 0.16 | (0.0, 0.0, 0.0) | (0.0, 0.0, 0.0) | −0.051 | 0.738 |
| SXS:BBH:0182 | 0.188 | (0.0, 0.0, 0.0) | (0.0, 0.0, 0.0) | −1.326 | −2.451 |
| SXS:BBH:0183 | 0.222 | (0.0, 0.0, 0.0) | (0.0, 0.0, 0.0) | 0.972 | −2.693 |
| SXS:BBH:0184 | 0.25 | (0.0, 0.0, 0.98) | (0.0, 0.0, 0.98) | −0.787 | −4.076 |
| SXS:BBH:0185 | 0.187 | (0.0, 0.0, 0.5) | (0.0, 0.0, 0.0) | 0.85 | −2.776 |
| SXS:BBH:0186 | 0.25 | (0.0, 0.0, 0.75) | (0.0, 0.0, 0.75) | −0.111 | 0.164 |
| SXS:BBH:0187 | 0.25 | (0.0, 0.0, 0.96) | (0.0, 0.0, 0.96) | −0.747 | 2.225 |
| SXS:BBH:0188 | 0.25 | (0.0, 0.0, 0.99) | (0.0, 0.0, 0.99) | −0.207 | −0.125 |
| SXS:BBH:0189 | 0.25 | (0.0, 0.0, 0.995) | (0.0, 0.0, 0.995) | 0.182 | −0.543 |
| SXS:BBH:0190 | 0.25 | (0.0, 0.0, 0.0) | (0.0, 0.0, 0.0) | 0.025 | 1.255 |
| SXS:BBH:0191 | 0.122 | (0.0, 0.0, 0.0) | (0.0, 0.0, 0.0) | 0.181 | 0.879 |
| SXS:BBH:0192 | 0.16 | (0.0, 0.0, 0.0) | (0.0, 0.0, 0.0) | −0.182 | 0.756 |
| SXS:BBH:0193 | 0.187 | (0.0, 0.0, 0.0) | (0.0, 0.0, 0.0) | 1.29 | 0.255 |
| SXS:BBH:0194 | 0.222 | (0.0, 0.0, 0.0) | (0.0, 0.0, 0.0) | −5.356 | −0.596 |
| SXS:BBH:0195 | 0.083 | (0.0, 0.0, 0.0) | (0.0, 0.0, 0.0) | 0.432 | −4.796 |
| SXS:BBH:0196 | 0.096 | (0.0, 0.0, 0.0) | (0.0, 0.0, 0.0) | −0.001 | 0.732 |
| SXS:BBH:0197 | 0.138 | (0.0, 0.0, 0.0) | (0.0, 0.0, 0.0) | −0.34 | 0.903 |
| SXS:BBH:0198 | 0.107 | (0.0, 0.0, 0.0) | (0.0, 0.0, 0.0) | −3.207 | −5.422 |
| SXS:BBH:0199 | 0.089 | (0.0, 0.0, 0.0) | (0.0, 0.0, 0.0) | −0.221 | 0.768 |
| SXS:BBH:0200 | 0.149 | (0.0, 0.0, 0.0) | (0.0, 0.0, 0.0) | 3.95 | 0.894 |
| SXS:BBH:0201 | 0.204 | (0.0, 0.0, 0.0) | (0.0, 0.0, 0.0) | 3.176 | 1.896 |
| SXS:BBH:0202 | 0.115 | (0.0, 0.0, 0.0) | (0.0, 0.0, 0.0) | −0.921 | 0.461 |
| SXS:BBH:0203 | 0.173 | (0.0, 0.0, 0.0) | (0.0, 0.0, 0.0) | −1.535 | 1.219 |
| SXS:BBH:0204 | 0.239 | (0.0, 0.0, 0.0) | (0.0, 0.0, 0.0) | −1.119 | 1.145 |
| SXS:BBH:0205 | 0.101 | (0.0, 0.0, 0.0) | (0.0, 0.0, 0.0) | 0.589 | 0.745 |
| SXS:BBH:0206 | 0.085 | (0.0, 0.0, 0.0) | (0.0, 0.0, 0.0) | −1.417 | 2.264 |
| SXS:BBH:0207 | 0.13 | (0.0, 0.0, 0.0) | (0.0, 0.0, 0.0) | −0.51 | 1.02 |
| SXS:BBH:0208 | 0.248 | (0.0, 0.0, 0.0) | (0.0, 0.0, 0.0) | −0.041 | −4.178 |
| SXS:BBH:0209 | 0.092 | (0.0, 0.0, 0.0) | (0.0, 0.0, 0.0) | 0.195 | −0.661 |
| SXS:BBH:0210 | 0.179 | (0.0, 0.0, 0.0) | (0.0, 0.0, 0.0) | 0.054 | −1.457 |
| SXS:BBH:0211 | 0.211 | (0.0, 0.0, 0.0) | (0.0, 0.0, 0.0) | 0.506 | −4.73 |
| SXS:BBH:0212 | 0.109 | (0.0, 0.0, 0.6) | (0.0, 0.0, 0.0) | 0.04 | −3.718 |
| SXS:BBH:0213 | 0.109 | (0.0, 0.0, 0.4) | (0.0, 0.0, 0.0) | 0.116 | −0.039 |
| SXS:BBH:0214 | 0.109 | (0.0, 0.0, 0.4) | (0.0, 0.0, 0.0) | 0.413 | 0.105 |
| SXS:BBH:0215 | 0.109 | (0.0, 0.0, −0.4) | (0.0, 0.0, 0.0) | 0.05 | 2.065 |
| SXS:BBH:0216 | 0.109 | (0.0, 0.0, −0.4) | (0.0, 0.0, 0.0) | 0.189 | 0.109 |
| SXS:BBH:0217 | 0.109 | (0.0, 0.0, −0.6) | (0.0, 0.0, 0.0) | 0.349 | −0.04 |
| SXS:BBH:0218 | 0.139 | (0.0, 0.0, −0.9) | (0.0, 0.0, 0.0) | 4.693 | −2.211 |
| SXS:BBH:0219 | 0.25 | (0.0, 0.0, −0.9) | (0.0, 0.0, −0.5) | 1.193 | 1.982 |
| SXS:BBH:0220 | 0.25 | (0.0, 0.0, −0.9) | (0.0, 0.0, 0.0) | −0.071 | 0.152 |
| SXS:BBH:0221 | 0.25 | (0.0, 0.0, −0.9) | (0.0, 0.0, 0.9) | 1.534 | −0.432 |
| SXS:BBH:0222 | 0.25 | (0.0, 0.0, −0.8) | (0.0, 0.0, −0.8) | −0.661 | 0.691 |
| SXS:BBH:0223 | 0.25 | (0.0, 0.0, −0.8) | (0.0, 0.0, 0.8) | 2.164 | 2.203 |

(Continued.)

**Table B1.** (Continued.)

| Waveform ID | $\eta$ | $\chi_1$ | $\chi_2$ | $\Delta v$ | $\Delta \theta$ |
|---|---|---|---|---|---|
| SXS:BBH:0224 | 0.25 | (0.0, 0.0, −0.625) | (0.0, 0.0, −0.25) | 0.351 | −0.409 |
| SXS:BBH:0225 | 0.25 | (0.0, 0.0, −0.6) | (0.0, 0.0, −0.6) | 0.323 | −0.597 |
| SXS:BBH:0226 | 0.25 | (0.0, 0.0, −0.6) | (0.0, 0.0, 0.0) | 0.344 | 0.083 |
| SXS:BBH:0227 | 0.25 | (0.0, 0.0, −0.6) | (0.0, 0.0, 0.6) | 0.352 | −0.531 |
| SXS:BBH:0228 | 0.25 | (0.0, 0.0, −0.5) | (0.0, 0.0, 0.5) | 0.034 | 1.084 |
| SXS:BBH:0229 | 0.25 | (0.0, 0.0, −0.5) | (0.0, 0.0, 0.9) | 0.298 | −0.545 |
| SXS:BBH:0230 | 0.25 | (0.0, 0.0, −0.4) | (0.0, 0.0, −0.8) | 0.111 | −0.624 |
| SXS:BBH:0231 | 0.25 | (0.0, 0.0, −0.4) | (0.0, 0.0, 0.8) | 1.494 | 1.56 |
| SXS:BBH:0232 | 0.25 | (0.0, 0.0, −0.3) | (0.0, 0.0, 0.0) | 1.849 | 1.817 |
| SXS:BBH:0233 | 0.25 | (0.0, 0.0, 0.3) | (0.0, 0.0, 0.0) | −0.449 | 1.556 |
| SXS:BBH:0234 | 0.25 | (0.0, 0.0, 0.4) | (0.0, 0.0, −0.8) | −0.784 | 3.297 |
| SXS:BBH:0235 | 0.25 | (0.0, 0.0, 0.4) | (0.0, 0.0, 0.8) | 1.262 | −1.979 |
| SXS:BBH:0236 | 0.25 | (0.0, 0.0, 0.5) | (0.0, 0.0, −0.9) | −1.284 | −1.948 |
| SXS:BBH:0237 | 0.25 | (0.0, 0.0, 0.6) | (0.0, 0.0, 0.0) | 0.294 | 2.663 |
| SXS:BBH:0238 | 0.25 | (0.0, 0.0, 0.6) | (0.0, 0.0, 0.6) | −4.591 | 2.405 |
| SXS:BBH:0239 | 0.25 | (0.0, 0.0, 0.65) | (0.0, 0.0, 0.25) | −0.344 | −0.832 |
| SXS:BBH:0240 | 0.25 | (0.0, 0.0, 0.8) | (0.0, 0.0, 0.8) | 33.268 | 0.223 |
| SXS:BBH:0241 | 0.25 | (0.0, 0.0, 0.9) | (0.0, 0.0, 0.0) | 9.188 | 0.164 |
| SXS:BBH:0242 | 0.25 | (0.0, 0.0, 0.9) | (0.0, 0.0, 0.5) | 30.183 | 0.36 |
| SXS:BBH:0243 | 0.222 | (0.0, 0.0, −0.871) | (0.0, 0.0, 0.85) | 0.93 | 2.039 |
| SXS:BBH:0244 | 0.222 | (0.0, 0.0, −0.85) | (0.0, 0.0, −0.85) | 0.759 | 2.489 |
| SXS:BBH:0245 | 0.222 | (0.0, 0.0, −0.6) | (0.0, 0.0, −0.6) | 21.9 | 0.48 |
| SXS:BBH:0246 | 0.222 | (0.0, 0.0, −0.6) | (0.0, 0.0, 0.0) | 3.116 | 0.946 |
| SXS:BBH:0247 | 0.222 | (0.0, 0.0, −0.6) | (0.0, 0.0, 0.6) | −1.425 | −4.179 |
| SXS:BBH:0248 | 0.222 | (0.0, 0.0, −0.5) | (0.0, 0.0, −0.5) | 4.404 | −2.246 |
| SXS:BBH:0249 | 0.222 | (0.0, 0.0, −0.371) | (0.0, 0.0, 0.85) | −15.913 | −5.173 |
| SXS:BBH:0250 | 0.222 | (0.0, 0.0, −0.3) | (0.0, 0.0, −0.3) | −6.795 | −5.292 |
| SXS:BBH:0251 | 0.222 | (0.0, 0.0, −0.3) | (0.0, 0.0, 0.0) | −13.571 | 0.98 |
| SXS:BBH:0252 | 0.222 | (0.0, 0.0, −0.3) | (0.0, 0.0, 0.3) | 0.442 | 0.045 |
| SXS:BBH:0253 | 0.222 | (0.0, 0.0, −0.129) | (0.0, 0.0, −0.85) | −0.564 | 1.978 |
| SXS:BBH:0254 | 0.222 | (0.0, 0.0, 0.0) | (0.0, 0.0, −0.6) | −1.315 | −2.272 |
| SXS:BBH:0255 | 0.222 | (0.0, 0.0, 0.0) | (0.0, 0.0, −0.3) | −1.016 | 2.421 |
| SXS:BBH:0256 | 0.222 | (0.0, 0.0, 0.0) | (0.0, 0.0, 0.3) | −1.275 | 1.997 |
| SXS:BBH:0257 | 0.222 | (0.0, 0.0, 0.0) | (0.0, 0.0, 0.6) | −4.439 | −2.458 |
| SXS:BBH:0258 | 0.222 | (0.0, 0.0, 0.129) | (0.0, 0.0, 0.85) | 0.101 | −3.043 |
| SXS:BBH:0259 | 0.222 | (0.0, 0.0, 0.3) | (0.0, 0.0, −0.3) | −7.843 | 2.171 |
| SXS:BBH:0260 | 0.222 | (0.0, 0.0, 0.3) | (0.0, 0.0, 0.0) | 0.889 | 4.094 |
| SXS:BBH:0261 | 0.222 | (0.0, 0.0, 0.3) | (0.0, 0.0, 0.3) | −0.433 | 3.291 |
| SXS:BBH:0262 | 0.222 | (0.0, 0.0, 0.371) | (0.0, 0.0, −0.85) | −0.219 | 2.466 |
| SXS:BBH:0263 | 0.222 | (0.0, 0.0, 0.5) | (0.0, 0.0, 0.5) | 1.253 | −3.211 |
| SXS:BBH:0264 | 0.222 | (0.0, 0.0, 0.6) | (0.0, 0.0, −0.6) | 0.459 | −2.289 |
| SXS:BBH:0265 | 0.222 | (0.0, 0.0, 0.6) | (0.0, 0.0, 0.0) | 0.215 | 0.47 |
| SXS:BBH:0266 | 0.222 | (0.0, 0.0, 0.6) | (0.0, 0.0, 0.6) | 0.501 | −1.989 |
| SXS:BBH:0267 | 0.222 | (0.0, 0.0, 0.85) | (0.0, 0.0, 0.85) | 1.319 | 0.058 |
| SXS:BBH:0268 | 0.222 | (0.0, 0.0, 0.871) | (0.0, 0.0, −0.85) | 0.541 | −0.085 |
| SXS:BBH:0269 | 0.204 | (0.0, 0.0, 0.0) | (0.0, 0.0, 0.0) | 0.286 | −1.487 |
| SXS:BBH:0270 | 0.188 | (0.0, 0.0, −0.85) | (0.0, 0.0, −0.85) | −7.625 | 0.114 |
| SXS:BBH:0271 | 0.188 | (0.0, 0.0, −0.731) | (0.0, 0.0, 0.85) | −13.344 | 0.157 |
| SXS:BBH:0272 | 0.187 | (0.0, 0.0, −0.6) | (0.0, 0.0, 0.0) | −4.21 | 0.256 |

(Continued.)

**Table B1.** (Continued.)

| | | | | | |
|---|---|---|---|---|---|
| SXS:BBH:0273 | 0.188 | (0.0, 0.0, −0.6) | (0.0, 0.0, 0.6) | 0.387 | 5.454 |
| SXS:BBH:0274 | 0.188 | (0.0, 0.0, −0.6) | (0.0, 0.0, −0.6) | 1.144 | 0.382 |
| SXS:BBH:0275 | 0.187 | (0.0, 0.0, −0.6) | (0.0, 0.0, −0.4) | −0.136 | 2.122 |
| SXS:BBH:0276 | 0.187 | (0.0, 0.0, −0.6) | (0.0, 0.0, 0.4) | −2.085 | −0.567 |
| SXS:BBH:0277 | 0.188 | (0.0, 0.0, −0.5) | (0.0, 0.0, −0.5) | 14.362 | −0.212 |
| SXS:BBH:0278 | 0.188 | (0.0, 0.0, −0.4) | (0.0, 0.0, −0.6) | −0.491 | 5.994 |
| SXS:BBH:0279 | 0.188 | (0.0, 0.0, −0.4) | (0.0, 0.0, 0.6) | −0.438 | −0.182 |
| SXS:BBH:0280 | 0.188 | (0.0, 0.0, −0.3) | (0.0, 0.0, −0.3) | −0.486 | −1.179 |
| SXS:BBH:0281 | 0.187 | (0.0, 0.0, −0.3) | (0.0, 0.0, 0.0) | 1.006 | 0.269 |
| SXS:BBH:0282 | 0.188 | (0.0, 0.0, −0.3) | (0.0, 0.0, 0.3) | 1.365 | 0.283 |
| SXS:BBH:0283 | 0.188 | (0.0, 0.0, −0.269) | (0.0, 0.0, −0.85) | 13.888 | 0.239 |
| SXS:BBH:0284 | 0.188 | (0.0, 0.0, −0.231) | (0.0, 0.0, 0.85) | 0.396 | 4.617 |
| SXS:BBH:0285 | 0.188 | (0.0, 0.0, 0.0) | (0.0, 0.0, −0.6) | 0.152 | 2.157 |
| SXS:BBH:0286 | 0.188 | (0.0, 0.0, 0.0) | (0.0, 0.0, −0.3) | −0.202 | 0.795 |
| SXS:BBH:0287 | 0.188 | (0.0, 0.0, 0.0) | (0.0, 0.0, 0.3) | −0.19 | −0.91 |
| SXS:BBH:0288 | 0.188 | (0.0, 0.0, 0.0) | (0.0, 0.0, 0.6) | −0.318 | 2.466 |
| SXS:BBH:0289 | 0.188 | (0.0, 0.0, 0.231) | (0.0, 0.0, −0.85) | −1.702 | 1.61 |
| SXS:BBH:0290 | 0.188 | (0.0, 0.0, 0.269) | (0.0, 0.0, 0.85) | −0.745 | −0.958 |
| SXS:BBH:0291 | 0.188 | (0.0, 0.0, 0.3) | (0.0, 0.0, −0.3) | −1.241 | 0.467 |
| SXS:BBH:0292 | 0.187 | (0.0, 0.0, 0.3) | (0.0, 0.0, 0.0) | −1.491 | −2.115 |
| SXS:BBH:0293 | 0.188 | (0.0, 0.0, 0.3) | (0.0, 0.0, 0.3) | 1.642 | −0.474 |
| SXS:BBH:0294 | 0.188 | (0.0, 0.0, 0.4) | (0.0, 0.0, −0.6) | 2.581 | 1.489 |
| SXS:BBH:0295 | 0.188 | (0.0, 0.0, 0.4) | (0.0, 0.0, 0.6) | 10.325 | −0.021 |
| SXS:BBH:0296 | 0.188 | (0.0, 0.0, 0.5) | (0.0, 0.0, 0.5) | −0.552 | 0.965 |
| SXS:BBH:0297 | 0.188 | (0.0, 0.0, 0.6) | (0.0, 0.0, −0.6) | 3.966 | −4.49 |
| SXS:BBH:0298 | 0.188 | (0.0, 0.0, 0.6) | (0.0, 0.0, −0.4) | 8.637 | 1.334 |
| SXS:BBH:0299 | 0.188 | (0.0, 0.0, 0.6) | (0.0, 0.0, 0.0) | 3.36 | −0.215 |
| SXS:BBH:0300 | 0.188 | (0.0, 0.0, 0.6) | (0.0, 0.0, 0.4) | 7.92 | −4.126 |
| SXS:BBH:0301 | 0.187 | (0.0, 0.0, 0.6) | (0.0, 0.0, 0.6) | 4.708 | 1.765 |
| SXS:BBH:0302 | 0.188 | (0.0, 0.0, 0.731) | (0.0, 0.0, −0.85) | 2.827 | 1.355 |
| SXS:BBH:0303 | 0.188 | (0.0, 0.0, 0.85) | (0.0, 0.0, 0.85) | 7.128 | −0.872 |
| SXS:BBH:0305 | 0.173 | (0.0, 0.0, 0.0) | (0.0, 0.0, 0.0) | 0.305 | 0.054 |
| SXS:BBH:0306 | 0.149 | (0.0, 0.0, 0.0) | (0.0, 0.0, 0.0) | 2.69 | −2.853 |
| SXS:BBH:0307 | 0.130 | (0.0, 0.0, 0.0) | (0.0, 0.0, 0.0) | −0.3 | 2.773 |
| SXS:BBH:0317 | 0.116 | (0.0, 0.0, 0.0) | (0.0, 0.0, 0.0) | −3.448 | 1.909 |
| SXS:BBH:0318 | 0.109 | (0.0, 0.0, 0.0) | (0.0, 0.0, 0.0) | 1.134 | −0.07 |
| SXS:BBH:0319 | 0.104 | (0.0, 0.0, 0.0) | (0.0, 0.0, 0.0) | −0.186 | −0.109 |
| SXS:BBH:0320 | 0.094 | (0.0, 0.0, 0.0) | (0.0, 0.0, 0.0) | −1.866 | −2.913 |

## Appendix C. NR error estimates as a function of the intrinsic parameters

We estimate the kick error in NR waveforms using SXS simulations which include at least two resolutions. Figures C1 and C2 show the errors as a function of the symmetric mass ratio and the spin components of the binary.

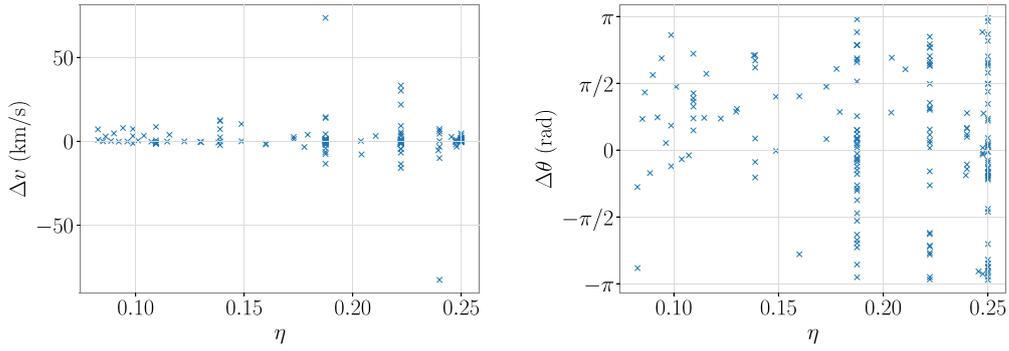

**Figure C1.** NR error estimates of the kick magnitude (left) and orientation (right) as a function of the symmetric mass ratio.

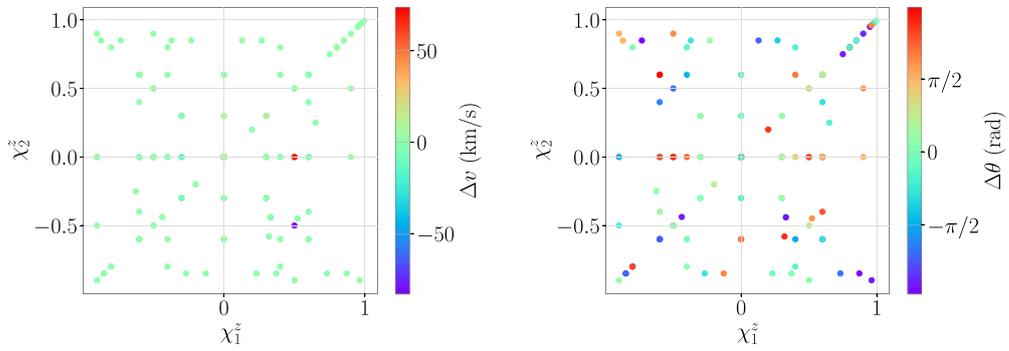

**Figure C2.** NR error estimates of the kick magnitude (left) and orientation (right) as a function of the individual spins.

## ORCID iDs

Angela Borchers ● https://orcid.org/0000-0002-2184-7388
Frank Ohme ● https://orcid.org/0000-0003-0493-5607